\documentclass[10pt,a4paper,pra,superscriptaddress,twocolumn,nofootinbib,floatfix,accepted=2025-09-15]{quantumarticle}
\pdfoutput=1

\usepackage[utf8]{inputenc}
\usepackage[english]{babel}
\usepackage[T1]{fontenc}
\usepackage{amsmath}

\usepackage{physics}
\usepackage{graphicx}
\usepackage{dcolumn}
\usepackage{bm}
\usepackage[breaklinks=true,colorlinks,citecolor=blue,linkcolor=blue,urlcolor=blue]{hyperref}
\usepackage{xcolor}

\newcommand{\figpanel}[2]{Fig.~\hyperref[#1]{\ref*{#1}(#2)}}

\usepackage[capitalise]{cleveref}

\usepackage[utf8]{inputenc}
\usepackage[T1]{fontenc}

\usepackage{multirow}

\usepackage{algorithm}
\usepackage{algpseudocode}

\definecolor{nred}{rgb}{0.9,0.1,0.1}
\definecolor{nblack}{rgb}{0,0,0}
\definecolor{nblue}{rgb}{0.2,0.2,0.8}
\definecolor{ngreen}{rgb}{0.2,0.6,0.2}

\usepackage{listings}
\definecolor{codegreen}{rgb}{0,0.6,0}
\definecolor{codegray}{rgb}{0.5,0.5,0.5}
\definecolor{codepurple}{rgb}{0.58,0,0.82}
\definecolor{codered}{rgb}{0.796,0.235,0.2}
\definecolor{backcolour}{rgb}{0.95,0.95,0.92}
\definecolor{dodgerblue4}{rgb}{0.06, 0.31, 0.55}

\lstdefinelanguage{Julia}{
    keywords=[1]{abstract, break, catch, const, continue, do, else, elseif, begin, end, export, false, for, function, if, import, let, local, macro, module, quote, return, struct, true, try, using, while, where},
    keywords=[2]{Bool, Char, Dict, Float64, Int, String, Array, AbstractArray, Vector, Set, Tuple, NamedTuple, QuantumObject, QuantumObjectEvolution},
    keywords=[3]{dummy_start, versioninfo, Qobj, QobjEvo, tensor, basis, fock, destroy, qeye, sigmax, sigmay, sigmaz, sigmam, sigmap, range, sesolve, mesolve, mcsolve, ssesolve, smesolve, sqrt, sin, cos, steadystate, steadystate_fourier, dfd_mesolve, dsf_mesolve, coherent, sum, size, dropdims, H_dsf, c_ops_dsf, e_ops_dsf, cu, CuVector, wigner, normalize, addprocs, SlurmManager, set_num_threads, Lattice, mapreduce, multisite_operator, DissipativeIsing, Val, ntuple, EnsembleSplitThreads, rmprocs, workers, gradient, my_f_mesolve, real, expect, BacksolveAdjoint, EnzymeVJP, liouvillian, exp, conj, coef_a, coef_ac, dummy_end},
    sensitive=true,  
    morecomment=[l]\#,%
    morecomment=[n]{\#=}{=\#},%
    morestring=[s]{"}{"},%
    literate={
        {\\}{{{\color{codered}\lstum@backslash}}}{1} {\{}{{{\color{codered}\{}}}{1}
        {\}}{{{\color{codered}\}}}}{1} 
        {\%}{{{\color{codered}\%}}}{1} {&}{{{\color{codered}\&}}}{1}
        {+}{{{\color{codered}+}}}{1}
        {*}{{{\color{codered}*}}}{1}
        {/}{{{\color{codered}/}}}{1}
        {'}{{{\color{codered}\textquotesingle}}}{1}
        {=}{{{\color{codered}=}}}{1}
        {<}{{{\color{codered}<}}}{1} 
        {=>}{{{\color{codered}=>}}}{1} {|>}{{{\color{codered}|>}}}{1}
        {==}{{{\color{codered}==}}}{1}
        {>}{{{\color{codered}>}}}{1} {?}{{{\color{codered}?}}}{1}
        {julia>}{{{\color{codered}julia>}}}{1}
        {^}{{{\color{codered}\textasciicircum}}}{1} {|}{{{\color{codered}|}}}{1}
        {~}{{{\color{codered}\textasciitilde{}}}}{1}
        {×}{{{\color{codered}$\times$}}}{1} {⋅}{{{\color{codered}$\cdot$}}}{1}
        {π}{{$\pi$}}1
        {ω}{{$\omega$}}1
        {α}{{$\alpha$}}1
        {β}{{$\beta$}}1
        {σ}{{$\sigma$}}1
        {ψ}{{$\psi$}}1
        {κ}{{$\kappa$}}1
        {γ}{{$\gamma$}}1
        {κ}{{$\kappa$}}1
        {Δ}{{$\Delta$}}1
        {φ}{{$\phi$}}1
        {∘}{{$\circ$}}1
    }
}

\lstdefinestyle{mainstyle}{
    language=Julia,
    backgroundcolor=\color{backcolour},   
    commentstyle=\color{gray},    
    keywordstyle=\color{blue}\bfseries,  
    keywordstyle=[2]\color{teal},  
    keywordstyle=[3]\color{dodgerblue4},  
    stringstyle=\color{codepurple},
    basicstyle=\ttfamily\footnotesize,
    columns=fullflexible,
    upquote=true,
    breakatwhitespace=false,         
    breaklines=true,                 
    captionpos=b,                    
    keepspaces=true,                 
    showspaces=false,                
    showstringspaces=false,
    showtabs=false,                  
    tabsize=2,
}

\lstdefinestyle{tablestyle}{
    language=Julia,
    backgroundcolor=\color{backcolour},
    keywordstyle=[2]\color{black},  
    keywordstyle=[3]\color{black},  
    basicstyle=\ttfamily\footnotesize,
    columns=fullflexible,
    upquote=true,
    breakatwhitespace=false,         
    breaklines=true,                 
    captionpos=b,                    
    keepspaces=true,                 
    numbers=left,                    
    numbersep=5pt,                  
    showspaces=false,                
    showstringspaces=false,
    showtabs=false,                  
    tabsize=2,
}

\newcommand{\code}[1]{%
  \begingroup
  \setlength{\fboxsep}{1pt} 
  \colorbox{backcolour}{\lstinline[basicstyle=\ttfamily]{#1}}%
  \endgroup
}
\newcommand{\codesymbol}[1]{%
  \begingroup
  \setlength{\fboxsep}{1pt} 
  \colorbox{backcolour}{#1}%
  \endgroup
}

\begin{document}

\title{QuantumToolbox.jl: An efficient Julia framework for simulating open quantum systems}

\author{Alberto Mercurio$^{*}$}
\affiliation{Institute of Physics, Ecole Polytechnique Fédérale de Lausanne (EPFL), CH-1015 Lausanne, Switzerland}
\affiliation{Center for Quantum Science and Engineering, Ecole Polytechnique Fédérale de Lausanne (EPFL), CH-1015 Lausanne, Switzerland}
\email{alberto.mercurio96@gmail.com}
\thanks{These authors contributed equally.}
\orcid{0000-0001-7814-1936}

\author{Yi-Te Huang$^{\dagger}$}
\affiliation{Department of Physics, National Cheng Kung University, Tainan 701401, Taiwan}
\affiliation{Center for Quantum Frontiers of Research and Technology (QFort), Tainan 701401, Taiwan}
\affiliation{RIKEN Center for Quantum Computing, RIKEN, Wakoshi, Saitama 351-0198, Japan}
\email{yitehuang.tw@gmail.com}
\thanks{These authors contributed equally.}
\orcid{0000-0002-2520-8348}

\author{Li-Xun Cai}
\orcid{0009-0006-7644-5200}
\affiliation{Department of Physics, National Cheng Kung University, Tainan 701401, Taiwan}
\affiliation{Center for Quantum Frontiers of Research and Technology (QFort), Tainan 701401, Taiwan}

\author{Yueh-Nan Chen}
\orcid{0000-0002-2785-7675}
\affiliation{Department of Physics, National Cheng Kung University, Tainan 701401, Taiwan}
\affiliation{Center for Quantum Frontiers of Research and Technology (QFort), Tainan 701401, Taiwan}
\affiliation{Physics Division, National Center for Theoretical Sciences, Taipei 106319, Taiwan}

\author{Vincenzo Savona}
\orcid{0000-0002-8984-6584}
\affiliation{Institute of Physics, Ecole Polytechnique Fédérale de Lausanne (EPFL), CH-1015 Lausanne, Switzerland}
\affiliation{Center for Quantum Science and Engineering, Ecole Polytechnique Fédérale de Lausanne (EPFL), CH-1015 Lausanne, Switzerland}

\author{Franco Nori}
\orcid{0000-0003-3682-7432}
\affiliation{RIKEN Center for Quantum Computing, RIKEN, Wakoshi, Saitama 351-0198, Japan}
\affiliation{Physics Department, The University of Michigan, Ann Arbor, Michigan 48109-1040, USA.}

\maketitle

\begin{abstract}
We present \texttt{QuantumToolbox.jl}, an open-source Julia package for simulating open quantum systems. Designed with a syntax familiar to users of \texttt{QuTiP} (Quantum Toolbox in Python), it harnesses Julia’s high-performance ecosystem to deliver fast and scalable simulations. The package includes a suite of time-evolution solvers supporting distributed computing and GPU acceleration, enabling efficient simulation of large-scale quantum systems. We also show how \texttt{QuantumToolbox.jl} can integrate with automatic differentiation tools, making it well-suited for gradient-based optimization tasks such as quantum optimal control. Benchmark comparisons demonstrate substantial performance gains over existing frameworks. With its flexible design and computational efficiency, \texttt{QuantumToolbox.jl} serves as a powerful tool for both theoretical studies and practical applications in quantum science.
\end{abstract}

~ 

\noindent GitHub: {\small\href{https://github.com/qutip/QuantumToolbox.jl}{https://github.com/qutip/QuantumToolbox.jl}}

\section{Introduction}

The simulation of quantum systems is a fundamental task in many areas of physics, including quantum optics, condensed matter physics, and quantum information science. The development of efficient numerical methods and software packages for simulating these systems has become increasingly important due to their growing complexity and the need for accurate predictions of their behavior. Although the simulation of these systems generally requires standard linear algebra operations, the computational complexity grows exponentially with the number of degrees of freedom. This makes it challenging to simulate large-scale quantum systems, especially when considering the effects of interactions with the environment. 

In many-body physics, approximate techniques render intractable problems accessible. Tensor network methods exploit locality of entanglement to efficiently represent wavefunctions~\cite{White1992Density,White1993Density_matrix,Vidal2003Efficient,Banuls2009Matrix,Schollwock2011DMRG,Orus2014A,Orus2019Tensor,Cirac2021Matrix,Fishman2022ITensor}. Quantum Monte Carlo utilizes stochastic sampling~\cite{Ceperley1986Quantum,Becca2017Quantum}, while the dynamical mean field theory approximate many-body systems as effective impurity problems~\cite{Georges1992Hubbard,Georges1996Dynamical}. Moreover, variational approaches~\cite{McMillan1965Ground,Ceperley1977MonteCarlo,Shi2018Variational,gravina2024adaptive,Wu2024Variational}, including neural quantum states~\cite{Carleo2017Solving,netket_2_2019,netket_3_2022}, have recently emerged as powerful methods. Additionally, quantum cumulant expansions~\cite{Kubo1962Generalized,Plankensteiner2022QuantumCumulants_jl} and phase space representations~\cite{Polkovnikov2010Phase} offer complementary insights into the quantum correlations and nonclassical features of the system.

Although these techniques have demonstrated great success, the simulation of quantum systems without approximation remains an important challenge. Therefore, exploring the computational efficiency of different implementations on different programming languages optimized for various hardware architectures is crucial. Since 1999, several software packages have been developed to this purpose, including \texttt{QOToolbox}~\cite{QOToolbox1999} in Matlab~\cite{Matlab2016}, \texttt{QuTiP}~\cite{QuTiP2012,QuTiP2013,QuTiP2022,QuTiP2024} and \texttt{dynamiqs}~\cite{dynamiqs2024} in Python~\cite{Python2009}, as well as \texttt{QuantumOptics.jl}~\cite{QuantumOptics2018} in Julia~\cite{Julia2012, Julia2017}.

A common limitation of most existing Python-based packages is their strict dependency on a specific numerical backend (e.g., \texttt{SciPy} or \texttt{JAX}). The addition of GPU and automatic differentiation support has required major restructuring of these code bases. Moreover, while the JAX backend provides excellent automatic differentiation, it restricts computations to JAX-compatible operations, limiting flexibility in data structures and control flows. These architectural choices imply that, if a new and superior numerical backend emerges in the future, adopting it would again require substantial refactoring of the code.

In this work, we introduce \texttt{QuantumToolbox.jl} (Quantum Toolbox in Julia), a fully open-source software package written in Julia~\cite{Julia2012, Julia2017}. The package has been designed from the ground up to overcome these structural limitations, providing a backend-agnostic and extensible platform for simulating open quantum systems. The package can be used with a syntax similar to that of \texttt{QuTiP} and relies on the Julia design philosophy: \textit{one can have machine performance without sacrificing human convenience}~\cite{Julia2012}. It combines the simplicity and efficiency of Julia with advanced features like distributed computing~\cite{Julia2017} and GPU acceleration~\cite{JuliaGPU2018}, making simulation of quantum systems more accessible and efficient. Furthermore, by wrapping some functions from other Julia packages~\cite{DifferentialEquations2017,LinearSolve2025}, we could further optimize the computational efficiency of simulations and achieve a significant speedup with respect to the corresponding methods in other packages~\cite{QuTiP2012,QuTiP2013,QuTiP2024,QuantumOptics2018,dynamiqs2024}.

Julia is a high-level, dynamic programming language designed for high-performance scientific computing. It was created in 2012 with the goal to combine the speed of C~\cite{C2006}, the flexibility of Ruby~\cite{Ruby2007}, the ease of use of Python~\cite{Python2009}, and the statistical and numerical computing capabilities of R~\cite{R2021} and Matlab~\cite{Matlab2016}. Its performance advantage comes from a just-in-time (JIT) compiler~\cite{LLVM2004}, which translates source code into machine code before execution, ensuring both efficiency and speed. Recently, Julia has gained increasing attention in research communities, particularly in fields requiring intensive computation, such as open quantum system dynamics~\cite{QuantumOptics2018,HOQST2022,Plankensteiner2022QuantumCumulants_jl,HierarchicalEOM2023,BundgaardNielsen2024Waveguideqed_jl}, quantum algorithms~\cite{Yao2020}, and quantum information~\cite{QuantumInformation2018}. Moreover, Julia supports UTF-8 variable names, allowing the usage of Greek and mathematical symbols, and thus, making the equations in code more readable and intuitive.

\begin{figure*}[ht]
    \centering \includegraphics[width=0.8\textwidth]{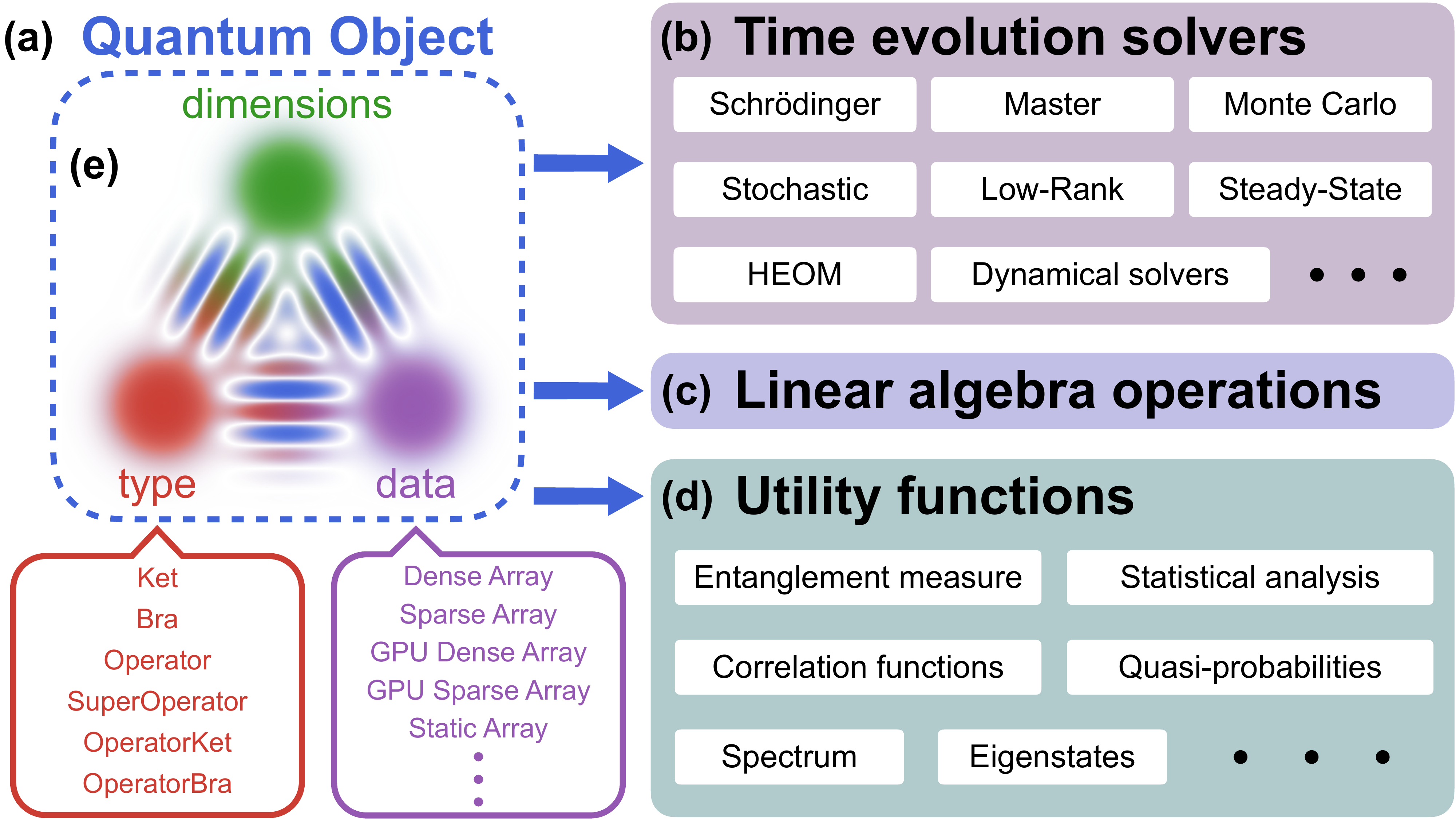}
    \caption{
        \textbf{Ecosystem of the \texttt{QuantumToolbox.jl} package.}
        \textbf{(a)} The quantum object structure encapsulates the properties of generic quantum objects through three main fields: \code{type}, \code{dimensions}, and \code{data}. The \code{type} field specifies the category of the quantum object, the \code{dimensions} field defines the Hilbert space structure for composite systems, and the \code{data} field supports arbitrary array types in Julia. \textbf{(b)} Various time evolution solvers are available for studying open quantum system dynamics. \textbf{(c)} All quantum objects support fundamental arithmetic and linear algebra operations. \textbf{(d)} \texttt{QuantumToolbox.jl} also provides a variety of utility functions for analyzing the properties and physical quantities of a given quantum object. \textbf{(e)} Logo of the \texttt{QuantumToolbox.jl} package.}
    \label{fig:QuantumToolbox.jl}
\end{figure*}

One can easily install \texttt{QuantumToolbox.jl} by running the following commands inside Julia's interactive session:
\begin{lstlisting}
import Pkg
Pkg.add("QuantumToolbox")
\end{lstlisting}
To load the package and check the version information, one can run the following commands:
\begin{lstlisting}
using QuantumToolbox
QuantumToolbox.versioninfo()
\end{lstlisting}
Note that all examples and code snippets presented in the following sections are tested with \texttt{QuantumToolbox.jl} version 0.34.1. The entire ecosystem of \texttt{QuantumToolbox.jl} package is summarized in Fig.~\ref{fig:QuantumToolbox.jl}.

\section{Package architecture and design philosophy.}
\label{sec-package-architecture}

In this Section, we summarize the main structure of the package and the most relevant user-accessible functions of \texttt{QuantumToolbox.jl}. A schematic is shown in Fig.~\ref{fig:QuantumToolbox.jl} and a complete list of available functions is shown in Appendix~\ref{app:table-of-functions}. By taking advantage of the Julia programming language features (e.g., multiple dispatch, metaprogramming, and distributed computing~\cite{Julia2012,Julia2017}) and other powerful Julia packages~\cite{JuliaGPU2018,DifferentialEquations2017,LinearSolve2025}, simulating large-scale open quantum system dynamics becomes very easy and efficient.

As shown in Fig.~\ref{fig:QuantumToolbox.jl}(a), the \code{QuantumObject} (or its abbreviated synonym \code{Qobj}) serves as a fundamental structure in \texttt{QuantumToolbox.jl}, representing quantum states, operators, and superoperators in a unified manner. The design of \code{Qobj} is inspired by similar abstractions found in \texttt{QuTiP} and also leverages Julia's multiple dispatch feature, allowing standard arithmetic operations and other advanced features for \code{Qobj}.

The \code{Qobj} structure is composed of three fields: \code{type}, \code{dimensions} (or \code{dims}), and \code{data}. The \code{type} field specifies whether a \code{Qobj} represents a quantum \code{Ket}, \code{Bra}, \code{Operator}, \code{SuperOperator}, \code{OperatorKet}, or \code{OperatorBra}, enabling intuitive manipulation of these objects. The \code{dimensions} (or \code{dims}) field stores the information about the Hilbert space of each subsystem in a composite system, ensuring compatibility with tensor products and partial traces on multipartite systems.

The \code{data} field contains the numerical representation of the quantum object. It stores the underlying data as a generic \code{AbstractArray}-type, allowing for broad compatibility with different computational backends. Depending on the nature of the quantum object and users' computational requirements, the \code{data} field can hold:
\begin{itemize}
    \item Standard Julia dense or sparse arrays
    \item GPU dense or sparse arrays~\cite{JuliaGPU2018}
    \item Statically sized arrays~\cite{StaticArrays_jl}
    \item Any other custom array that conforms to Julia's \code{AbstractArray} type.
\end{itemize}
This feature makes the \code{Qobj} highly adaptable across different platforms and architectures, allowing efficient large-scale quantum simulations without modifying the high-level application programming interface (API).

The \code{QuantumObjectEvolution} (or its abbreviated synonym \code{QobjEvo}) structure is introduced to encode time-dependent quantum objects for time evolution simulations. It shares the same fields as \code{Qobj}, with the \code{data} field leveraging the package \texttt{SciMLOperators.jl}~\cite{SciMLOperators_jl}, which is one of the core packages of \texttt{DifferentialEquations.jl}~\cite{DifferentialEquations2017}.

We demonstrate the usage of \code{Qobj} and \code{QobjEvo} by constructing the following time-dependent Hamiltonian for a single spin-1/2 system:
\begin{equation}
    \begin{split}
        \hat{H}(t) =& \ 
        \begin{pmatrix}
            1 &  0\\
            0 & -1
        \end{pmatrix} + \cos(\omega_1 t)
        \begin{pmatrix}
            0 & 1\\
            1 & 0
        \end{pmatrix} \\
        &+ \sin(\omega_2 t)
        \begin{pmatrix}
            0 & -i\\
            i &  0
        \end{pmatrix},
    \end{split}
\end{equation}
This Hamiltonian can be generated by the following code:
\begin{lstlisting}
# time-dependent coefficient functions
## parameters passed as Vector
p = [ω1, ω2]
f1(p, t) = cos(p[1] * t)
f2(p, t) = sin(p[2] * t)

## or

## parameters passed as NamedTuple
p = (ω1 = ω1, ω2 = ω2)
f1(p, t) = cos(p.ω1 * t)
f2(p, t) = sin(p.ω2 * t)

# Pauli matrices
## same as calling sigmax(), sigmay(), and sigmaz()
σx=Qobj([0  1.0;   1.0   0], type=Operator(), dims=2)
σy=Qobj([0 -1.0im; 1.0im 0], type=Operator(), dims=2)
σz=Qobj([1.0  0;   0  -1.0], type=Operator(), dims=2)

H = QobjEvo((σz, (σx, f1), (σy, f2)))

H(p, 0.1)  # returns the Hamiltonian at t = 0.1
\end{lstlisting}
A time-dependent quantum object can be built by calling the function \code{QobjEvo} and providing one or more time-independent \code{Qobj} along with corresponding time-dependent coefficient functions. The type of parameter \code{p} for the coefficient function can be either a \code{Vector} or \code{NamedTuple}. Notice that the Pauli matrices in this example can be more simply constructed by calling the functions \code{sigmax()}, \code{sigmay()}, and \code{sigmaz()}. \texttt{QuantumToolbox.jl} provides many functions for constructing typical quantum states and operators, as listed in Table~\ref{tab:Func-create-Qobj}.

Figure~\ref{fig:QuantumToolbox.jl}(b) shows that \texttt{QuantumToolbox.jl} can solve a wide range of time evolution problems, including deterministic and stochastic dynamics, variational low-rank~\cite{gravina2024adaptive} and hierarchical equations of motion (HEOM)~\cite{HierarchicalEOM2023,tanimura1989,tanimura1990,tanimura2020,jin2008,li2012,QuTiP-BoFiN2023,kuo2023,kuo2025}. A summary of the available time evolution solvers is provided in Table~\ref{tab:Func-Solvers}.

By utilizing the \texttt{DifferentialEquations.jl}~\cite{DifferentialEquations2017} package, which provides a set of low-level solvers for ordinary and stochastic differential equations, \texttt{QuantumToolbox.jl} achieves enhanced performance and scalability for time evolution simulations. To compute stationary states, we leverage the LinearSolve.jl~\cite{LinearSolve2025} package, which provides a unified and efficient interface to solve linear equations using various algorithms. We emphasize that solving the aforementioned problems on GPUs is straightforward by specifying the \code{data} field of \code{Qobj} as a GPU array~\cite{JuliaGPU2018}. Moreover, the stochastic (or quantum trajectory) solvers support distributed computing~\cite{Julia2017,DifferentialEquations2017}, enabling parallel integration of many trajectories on large-scale systems.

As can be seen from Fig.~\ref{fig:QuantumToolbox.jl}(b), HEOM can be used to study systems strongly coupled to the environment using the \texttt{HierarchicalEOM.jl} package~\cite{HierarchicalEOM2023}. This package is built on top of \texttt{QuantumToolbox.jl} and provides a user-friendly framework for simulating open quantum systems based on the HEOM approach~\cite{tanimura1989,tanimura1990,tanimura2020,jin2008,li2012,QuTiP-BoFiN2023,kuo2023,kuo2025}. The HEOM method enables a numerically exact characterization of all environmental effects on the system. It assumes that the environment is Gaussian and that the environmental operator in the interaction Hamiltonian couples linearly to the system, thereby preserving Gaussian statistics throughout the system-environment dynamics~\cite{tanimura1989,tanimura1990,tanimura2020}. For a detailed explanation of \texttt{HierarchicalEOM.jl} package, we refer the readers to Ref.~\cite{HierarchicalEOM2023}.

As shown in Fig.~\ref{fig:QuantumToolbox.jl}, \texttt{QuantumToolbox.jl} overloads most of the linear algebra operations provided by Julia's standard library \texttt{LinearAlgebra.jl}~\cite{Julia2012}. It also offers a variety of utility functions for analyzing the properties and physical quantities of \code{Qobj}, including the calculation of expectation values, eigen decompositions, quasi-probabilities, entanglement measurements, correlation functions, spectra, and more. The most relevant linear algebra and utility functions supported by \texttt{QuantumToolbox.jl} are summarized in Table~\ref{tab:Func-LA-Utility}.

Fig.~\ref{fig:QuantumToolbox.jl}(e) shows the logo of \texttt{QuantumToolbox.jl}, which is generated by the package itself. The logo represents the Wigner function~\cite{Wigner1932On} of a triangular cat state subject to decoherence. The state is constructed as a linear superposition of three coherent states using the \code{coherent} function, normalized with the \code{normalize} function, and evolved under the Lindblad master equation using the \code{mesolve} function, which we will discuss in \cref{sec-mesolve}. The Wigner quasi-probability is then computed using the \code{wigner} function.

In the following sections, we will demonstrate the features and capabilities of \texttt{QuantumToolbox.jl} with several examples. All figures are generated using the \texttt{Makie.jl}~\cite{DanischKrumbiegel2021} plotting library, written purely in Julia. The code used to generate the figures is available in a dedicated repository~\cite{QuantumToolbox_jl_Figures}.

\section{Time evolution solvers}

In this section, we demonstrate the features and capability of \texttt{QuantumToolbox.jl} with several examples.

\subsection{Schr\"odinger equation}
Given a quantum state $|\psi(0)\rangle$, the corresponding time-evolved state $|\psi(t)\rangle$ is given by the Schrödinger equation ($\hbar$ is set to unity throughout this work)
\begin{equation}
    \label{eq-schrodinger}
    i\frac{d}{dt}|\psi(t)\rangle = \hat{H} |\psi(t)\rangle \, ,
\end{equation}
where $\hat{H}$ is the Hamiltonian operator. Aside from a few solutions that can be found analytically, the Schr\"odinger equation is usually solved numerically. Thanks to the \code{sesolve} function, \texttt{QuantumToolbox.jl} provides a simple way to solve the Schr\"odinger equation for a given Hamiltonian and initial state.

As a pedagogical example, let us consider the Jaynes-Cummings (JC) model \cite{Jaynes1963Comparison}, which describes the interaction between a two-level system and a mode of the electromagnetic field. The Hamiltonian of the JC model is given by

\begin{equation}
    \label{eq-JC-hamiltonian}
    \hat{H} = \omega_c \hat{a}^\dagger \hat{a} + \frac{\omega_a}{2} \hat{\sigma}_z + g (\hat{a} \hat{\sigma}_+ + \hat{a}^\dagger \hat{\sigma}_-) \,
\end{equation}
where $\omega_c$ and $\omega_a$ are the frequencies of the cavity and the atom, respectively, $g$ is the coupling strength, and $\hat{a}$, $\hat{a}^\dagger$, and $\hat{\sigma}_j$ are the annihilation, creation and Pauli operators, respectively.

Consider the case where $\omega_c = \omega_a = 1$, $g = 0.1$, and the initial state is $|\psi(0)\rangle=|0\rangle\otimes|g\rangle$, where $|0\rangle$ is the vacuum state of the cavity and $|g\rangle$ is the ground state of the two-level system. In the following, we will solve the Schr\"odinger equation for this system and compute the expectation value of the number operator $\langle \hat{a}^\dagger \hat{a} \rangle$ as a function of time. We first define the parameters and the Hamiltonian of the JC model.
\begin{lstlisting}
ωc = 1
ωa = 1
g = 0.1

# cavity
N = 10 # Cutoff of the cavity Hilbert space
a = tensor(destroy(N), qeye(2)) #annihilation operator

# atom
σz = tensor(qeye(N), sigmaz()) # Pauli-Z operator
σm = tensor(qeye(N), sigmam()) # annihilation operator
σp = tensor(qeye(N), sigmap()) # creation operator

H = ωc * a' * a + ωa/2 * σz + g * (a * σp + a' * σm)
\end{lstlisting}
The time evolution can be simply simulated by
\begin{lstlisting}
# Initial cavity vacuum state and atomic excited state
ψ0 = tensor(fock(N, 0), basis(2, 0)) 

e_ops = [a' * a] # average photon number

tlist = range(0, 10*π/g, 1000)
sol_se = sesolve(H, ψ0, tlist, e_ops=e_ops)
sol_se.expect # get expectation values
\end{lstlisting}
Note how the syntax is mostly similar to the one of \texttt{QuTiP}. The expectation value $\langle\hat{a}^\dagger\hat{a}\rangle$ at each time point in \code{tlist} can be extracted by \code{sol_se.expect[1,:]}.

\begin{figure*}[t]
    \centering
    \includegraphics[width=\textwidth]{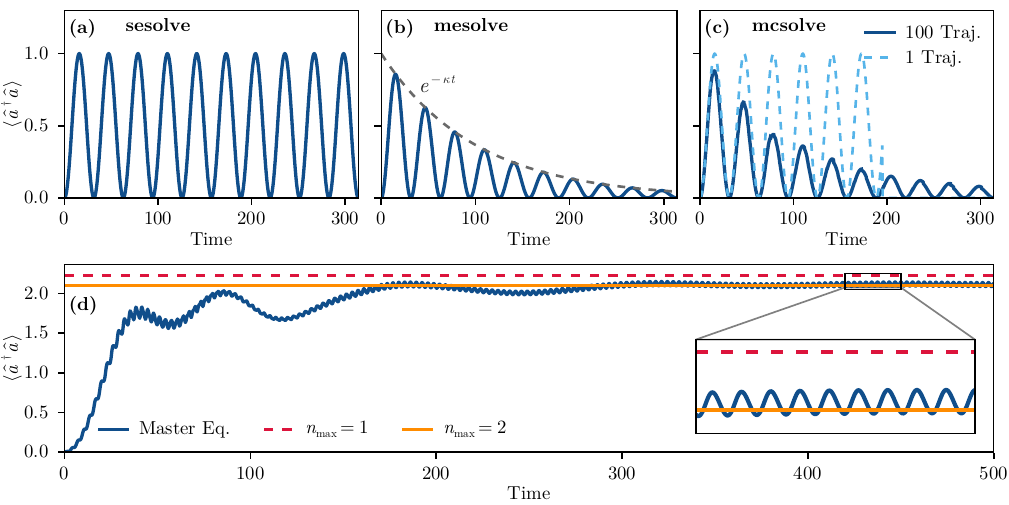}
    \caption{\textbf{Examples of time evolution solvers.} We observe the average photon number of the cavity $\langle \hat{a}^\dagger \hat{a} \rangle$ with respect to time in the Jaynes-Cummings model (\textbf{(a)}, \textbf{(b)} and \textbf{(c)}) and in the optomechanical system (\textbf{(d)}). \textbf{(a)} shows the time evolution following the Schr\"odinger equation, while \textbf{(b)} and \textbf{(c)} show the time evolution of the Lindblad master equation and the Monte Carlo wave-function, respectively. The dashed light blue curve in \textbf{(c)} represents a single quantum trajectory, where a quantum jump occurs at $t \approx 200$, and the solid blue curve represents the average over 100 quantum trajectories. \textbf{(d)} shows the time evolution of the optomechanical system, where the cavity is driven by a time-dependent drive. The dashed red and solid orange curves represent the value of the cavity population at the time-averaged steady state using $n_\mathrm{max}=1$ and $n_\mathrm{max}=2$ fourier components, respectively.}
    \label{fig-se-me-mc}
\end{figure*}

As can be seen from \cref{fig-se-me-mc}(a), the number of photons oscillates between the cavity and the atom, showing the vacuum Rabi oscillations. Its is important to note that the memory allocated does not depend on the number of points in the time grid or the final time. This is because the solver uses the most efficient algorithms given by the \texttt{DifferentialEquations.jl} package~\cite{DifferentialEquations2017}, computing intermediate steps in-place. This method only allocates extra memory if the user requests storage of the states with the \code{saveat} keyword argument, specifying the times at which the states are saved. Note that, when \code{e\_ops} is not defined, \code{saveat=tlist}.

\subsection{Lindblad master equation}\label{sec-mesolve}
The Schr\"odinger equation is a unitary evolution equation, which describes the dynamics of a closed quantum system. However, in many cases, the system is not isolated as it interacts with an environment. In the case of a Markovian environment, the dynamics of the system is described by the Lindblad master equation~\cite{Breuer2007The}
\begin{equation}
    \label{eq-lindblad-master-equation}
    \frac{d}{dt}\hat{\rho} = \mathcal{L} \hat{\rho} = -i [\hat{H}, \hat{\rho}] + \sum_k \mathcal{D}[\hat{C}_k] \hat{\rho} \, ,
\end{equation}
where $\mathcal{L}$ is the Liouvillian superoperator, $\hat{\rho}$ is the density matrix of the system, $\hat{H}$ is the Hamiltonian operator, $\hat{C}_k$ are the collapse operators, and 
\begin{equation}
    \mathcal{D}[\hat{C}] \hat{\rho} = \hat{C} \hat{\rho} \hat{C}^\dagger - \frac{1}{2} \left( \hat{C}^\dagger \hat{C} \hat{\rho} - \hat{\rho} \hat{C}^\dagger \hat{C} \right)
\end{equation}
is the Lindblad dissipator. \texttt{QuantumToolbox.jl} provides the \code{mesolve} function to solve the master equation for a given Hamiltonian, initial state, and collapse operators.

As an example, we consider the Jaynes-Cummings model, where both the cavity and the atom are coupled to a zero-temperature bath. The collapse operators are given by
\begin{equation}
    \hat{C}_1 = \sqrt{\kappa} \hat{a} \, , \quad \hat{C}_2 = \sqrt{\gamma} \hat{\sigma}_- \, ,
\end{equation}
where $\kappa$ and $\gamma$ are the decay rates of the cavity and the atom, respectively. We solve the master equation for this system and compute the expectation value of the number operator $\langle \hat{a}^\dagger \hat{a} \rangle$ as a function of time.
\begin{lstlisting}
κ = 0.01
γ = 0.01

C1 = sqrt(κ) * a
C2 = sqrt(γ) * σm

c_ops = [C1, C2]

sol_me = mesolve(H, ψ0, tlist, c_ops, e_ops=e_ops)
sol_me.expect # get expectation values
\end{lstlisting}
The expectation value $\langle\hat{a}^\dagger\hat{a}\rangle$ at each time point in \code{tlist} can be extracted by \code{sol_me.expect[1,:]}.

Figure~\ref{fig-se-me-mc}(b) shows the expectation value of the number operator $\langle \hat{a}^\dagger \hat{a} \rangle$ as a function of time. Contrary to the Schr\"odinger equation, the cavity population decays exponentially due to the cavity decay rate $\kappa$. Indeed, the dashed gray curve represents the exponential decay of the cavity population due to the cavity decay rate $\kappa$. The master equation solver is able to capture this behavior, as well as the oscillations due to the atom-cavity coupling.

When only the steady state is of interest, the \code{steadystate} function can be used to directly compute the steady state, without the need to solve the master equation in time. This function supports multiple algorithms, including direct, iterative, and eigenvalue-based methods~\cite{Nation2015Iterative,Nation2015Steady_state}.

\subsection{Monte Carlo wave-function}
The Monte Carlo quantum trajectories' approach is a stochastic method to solve the time evolution of an open quantum system~\cite{Dalibard1992Wave_function, Dum1992Monte, Molmer1993Monte, Carmichael1993An}. The idea is to simulate the evolution of the system by sampling the trajectories of the quantum state through a non-unitary evolution of a pure state $|\psi(t)\rangle$. The density matrix $\hat{\rho}(t)=\overline{|\psi(t)\rangle\langle\psi(t)|}$ obtained by averaging over the statistical ensemble of the trajectories is then an estimate of the actual solution of the Lindblad Master equation. The Monte Carlo method is particularly useful when the Hilbert space of the system is large, as it avoids direct computation of the density matrix and the Liouvillian superoperator. Trajectories can be easily computed in parallel, especially when using the Distributed.jl package in Julia. \texttt{QuantumToolbox.jl} provides the \code{mcsolve} function to solve the open quantum system dynamics using the Monte Carlo method. The system evolves according to the Schr\"odinger equation expressed in \cref{eq-schrodinger}, but with the non-Hermitian effective Hamiltonian
\begin{equation}
    \hat{H}_\mathrm{eff} = \hat{H} - \frac{i}{2} \sum_k \hat{C}_k^\dagger \hat{C}_k \, .
\end{equation}
Being the evolution non-unitary, the norm of the state is not conserved. A quantum jump occurs when the norm of the state reaches a random threshold, previously sampled from a uniform distribution between 0 and 1. If many jump operators are present, the jump to be executed is sampled from the probability distribution
\begin{equation}
    P_k (t) = \frac{\langle \psi(t) | \hat{C}_k^\dagger \hat{C}_k | \psi(t) \rangle}{\sum_j \langle \psi(t) | \hat{C}_j^\dagger \hat{C}_j | \psi(t) \rangle} \, .
\end{equation}
We consider the same system as before, where both the cavity and the atom are coupled to a zero-temperature bath. In addition to other keyword arguments, here we also use the \code{ntraj} and \code{rng} arguments to specify the number of quantum trajectories and the random seed, respectively. The second argument is optional, but it is useful for reproducibility.
\begin{lstlisting}
using Random
rng = MersenneTwister(10)

sol_mc = mcsolve(H, ψ0, tlist, c_ops, e_ops=e_ops, ntraj=100, rng=rng)
sol_mc.expect # get expectation values
\end{lstlisting}
The average expectation value $\langle\hat{a}^\dagger\hat{a}\rangle$ at each time point in \code{tlist} can be extracted by \code{sol_mc.expect[1,:]}.

\cref{fig-se-me-mc}(c) shows the expectation value of the number operator $\langle \hat{a}^\dagger \hat{a} \rangle$ as a function of time. As can be seen, by averaging over many trajectories, we get the same results as the master equation.

\subsection{Time-dependent Lindblad master equation}

In many models of interest, the Hamiltonian and-or the collapse operators are time-dependent. For example, this can happen when a drive is applied to the system or when dynamically changing the resonance frequency. Here, we show how the solvers can be used to simulate systems with time-dependent Hamiltonian or collapse operators. As an example, we consider the driven optomechanical system~\cite{Law1995Interaction,Aspelmeyer2014Cavity,Macri2018Nonperturbative}
\begin{equation}
\label{eq-optomechanical-driven}
\begin{split}
    \hat{H} =& \ \omega_c \hat{a}^\dagger \hat{a} + \omega_m \hat{b}^\dagger \hat{b} + \frac{g}{2} \left(\hat{a} + \hat{a}^\dagger\right)^2 \left(\hat{b} + \hat{b}^\dagger\right)  \\
    &+ F \cos \left(\omega_\mathrm{d} t \right) \left( \hat{a} + \hat{a}^\dagger \right) \, ,
\end{split}
\end{equation}
where $\hat{a}$ and $\hat{b}$ are the annihilation operators of the cavity and mechanical mode, respectively. The parameter $\omega_m$ represents the mechanical frequency, while $F$ and $\omega_\mathrm{d}$ denote the amplitude and frequency of the drive, respectively. It is worth noting that the time-dependence can not be removed by a unitary transformation, as the Hamiltonian contains counter-rotating terms on both the coupling and the drive. The time evolution is solved as before, with Hamiltonian being either a \code{Tuple} or a \code{QuantumObjectEvolution} type. Finally, we solve the open dynamics of the system by considering the cavity decay and the mechanical damping, with the collapse operators $\mathcal{D} [\hat{a}]$ and $\mathcal{D} [\hat{b}]$, respectively.

\begin{lstlisting}
Nc = 10 # Cutoff of the cavity Hilbert space
Nm = 7  # Cutoff of the mechanical mode Hilbert space
ωc = 1
ωm = 2 * ωc
g = 0.05
κ = 0.01
γ = 0.01
F = 10*κ
ωd = ωc

# time-dependent coefficient function
coef(p, t) = p[1] * cos(p[2] * t)

# annihilation operators
a = tensor(destroy(Nc), qeye(Nm)) # cavity
b = tensor(qeye(Nc), destroy(Nm)) # mechanical mode

H = ωc * a' * a + ωm * b' * b + 
    g / 2 * (a + a')^2 * (b + b')
c_ops = [sqrt(κ) * a, sqrt(γ) * b]
e_ops = [a' * a]
H_td = (H, (a + a', coef))

# Zero bare cavity and mechanical excitations
ψ0 = tensor(fock(Nc, 0), fock(Nm, 0))

params = [F, ωd]

tlist2 = range(0, 5/κ, 5000)
sol_me_td = mesolve(H_td, ψ0, tlist2, c_ops, e_ops=e_ops, params=params)
sol_me_td.expect # get expectation values
\end{lstlisting}
Note that, as we mentioned in Sec.~\ref{sec-package-architecture}, the \code{params} argument is also supported as a \code{NamedTuple} instead of a \code{Vector}. The evolution of the expectation value $\langle \hat{a}^\dagger \hat{a} \rangle$, extracted by \code{sol_me_td.expect[1,:]}, is shown in \cref{fig-se-me-mc}(d), where we can observe a stroboscopic behavior of the steady state due to the counter-rotating terms on both the coupling and the drive. This results in the impossibility of using the \code{steadystate} function to compute the steady state. To this end, \texttt{QuantumToolbox.jl} offers a solution by providing the \code{steadystate\_fourier} function. The method extracts the time-averaged steady state by expanding it into Fourier components. Depending on the internal solver, it then solves a matrix continuation fraction problem or a linear system~\cite{majumdar2011probing,papageorge2012bichromatic,maragkou2013bichromatic,Macri2022Spontaneous}. The case of the linear system solving approach is discussed in Appendix~\ref{app:steadystate-for-TD-systems}.

Although we applied this method to the master equation, the same approach can be used for all the time evolution solvers of \texttt{QuantumToolbox.jl}.

\subsection{Stochastic processes under continuous measurements}

Continuous quantum measurements such as homodyne and heterodyne detection describe the gradual acquisition of information about a quantum system in real time, leading to stochastic dynamics governed by measurement backaction~\cite{Wiseman2009Quantum}. These processes are fundamental for modeling measurement-based state estimation, e.g., the qubit readout~\cite{DiVincenzo2000Physical,Krantz2019A,buluta2009,georgescu2014}. In \texttt{QuantumToolbox.jl}, we have implemented the stochastic evolution of open quantum systems under homodyne detection, where a single quadrature of the field is continuously monitored. This corresponds to solving stochastic Schrödinger or master equations driven by Wiener noise, following the formalism used in \texttt{QuTiP}. The homodyne measurement allows access to real-time information about the system and is widely used in cavity and circuit QED experiments.

\subsubsection{Stochastic Schr\"odinger equation}

The stochastic Schr\"odinger equation (SSE) describes the time evolution of a quantum system under homodyne detection. Following this protocol, a detector does not resolve single photons but rather produces a homodyne current $J_x$~\cite{Wiseman2009Quantum}. The SSE is defined in Itô form by the following stochastic differential equation~\cite{Wiseman2009Quantum}

\begin{equation}
    d|\psi(t)\rangle = -i \hat{K} |\psi(t)\rangle dt + \sum_n \hat{M}_n |\psi(t)\rangle dW_n(t) \, ,
\end{equation}
where
\begin{align}
    \hat{K} &= \hat{H} + i \sum_n \left(\frac{e_n}{2} \hat{S}_n - \frac{1}{2} \hat{S}_n^\dagger \hat{S}_n - \frac{e_n^2}{8}\right) \, , \\
    \hat{M}_n &= \hat{S}_n - \frac{e_n}{2} \, , \\
    e_n &= \langle \psi(t) | \hat{S}_n + \hat{S}_n^\dagger | \psi(t) \rangle \, .
\end{align}
Here, $\hat{H}$ is the Hamiltonian, $\hat{S}_n$ are the stochastic collapse operators, and  $dW_n(t)$ is the real Wiener increment (associated to $\hat{S}_n$) with the expectation values of $E[dW_n]=0$ and $E[dW_n^2]=dt$, with $dt$ being the infinitesimal time step.

\begin{figure}[t]
    \centering
    \includegraphics{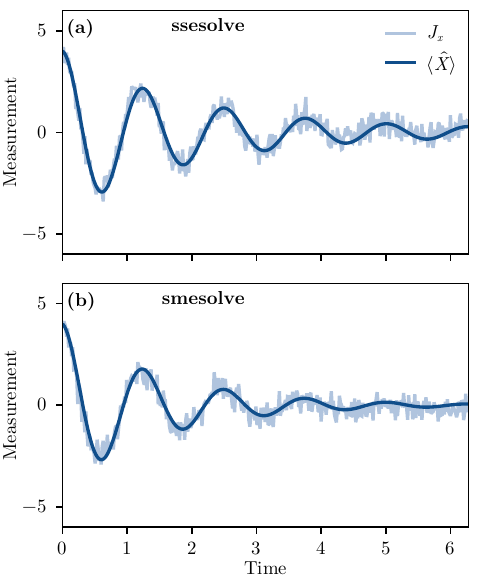}
    \caption{\textbf{Stochastic Schr\"odinger equation and stochastic master equation.} \textbf{(a)} Time evolution under the stochastic Schr\"odinger equation of the homodyne current $J_x$ (light blue curve) and the expectation value $\langle \hat{X} \rangle$ (dark blue curve) for the Jaynes-Cummings model. \textbf{(b)} Time evolution under the stochastic master equation of the same quantities as in \textbf{(a)}. Both the homodyne current and the expectation value are averaged over 500 trajectories.}
    \label{fig-sse-sme}
\end{figure}

As an example, we consider the JC Hamiltonian defined in \cref{eq-JC-hamiltonian}, where we perform homodyne detection on the cavity field $\hat{S} = \sqrt{\kappa} \hat{a}$. This protocol gives us access to the quadrature $\hat{X} = \hat{S} + \hat{S}^\dagger$ of the cavity field. We initialize the system with the cavity in a coherent state and the qubit in the excited state, $\ket{\psi (0)} = \ket{\alpha, e}$. The solver \code{ssesolve} will construct the operators $\hat{K}$ and $\hat{M}_n$, once the user passes the Hamiltonian \code{H} and the stochastic collapse operators list \code{sc_ops}. The homodyne current $J_x$ can be computed using~\cite{Wiseman2009Quantum}

\begin{equation}
    \label{eq-homodyne-current}
    J_x = \langle \hat{X} \rangle + \frac{dW}{dt} \, .
\end{equation}

\begin{lstlisting}
N = 20      # Fock space dimension
ωc = 5      # cavity resonance frequency
ωq = 5      # qubit resonance frequency
g = 0.1     # coupling strength
κ = 1       # cavity decay rate
γ = 2       # qubit decay rate
α = 2       # coherence of initial state
ntraj = 500 # number of trajectories

tlist = range(0, 10π / ωc, 400)

# operators
a = tensor(destroy(N), qeye(2))
σm = tensor(qeye(N), sigmam())
σz = tensor(qeye(N), sigmaz())

H = ωc * a' * a + ωq * σz / 2 +
    g * (a' * σm + a * σm')

# initial state |α, e>
ψ0 = tensor(coherent(N, α), basis(2, 0))

# stochastic collapse operators
sc_ops = [sqrt(κ) * a]

X = (a + a') * sqrt(κ)

sol_sse = ssesolve(
    H,
    ψ0,
    tlist,
    sc_ops[1],
    e_ops = [X],
    ntraj = ntraj,
    store_measurement = Val(true),
)
sol_sse.expect      # get expectation values
sol_sse.measurement # get homodyne current
\end{lstlisting}

The averaged expectation value $\langle \hat{X} \rangle$ can be extracted as in the previous examples with \code{sol_sse.expect}, while the homodyne current can be obtained using \code{sol_sse.measurement}. Figure~\ref{fig-sse-sme}(a) shows the time evolution of the homodyne current $J_x$ (light blue curve) and the expectation value $\langle \hat{X} \rangle$ (dark blue curve), averaged over 500 trajectories. It is worth mentioning that if \code{sc_ops} is a single \code{QuantumObject} rather than a \code{Vector} or a \code{Tuple}, the \code{ssesolve} function will use a different solver, specific for a single Wiener process. Indeed, when giving a list of operators, the solver uses an algorithm supporting multiple Wiener processes (non-diagonal noise). It is recommended to insert a single operator, when applicable, as this improves both the performance and the stability of the solver.

\subsubsection{Stochastic master equation}
The stochastic master equation (SME) describes the evolution of the density matrix of a quantum system under continuous measurements. It can also take into account losses and dephasing not related to the measurement process, or inefficient measurement channels. The stochastic master equation in the Itô form is given by~\cite{Wiseman2009Quantum}

\begin{equation}
    \begin{split}
        d \rho (t) = & \ -i [\hat{H}, \rho(t)] dt + \sum_i \mathcal{D}[\hat{C}_i] \rho(t) dt \\
        &+ \sum_n \mathcal{D}[\hat{S}_n] \rho(t) dt \\
        &+ \sum_n \mathcal{H}[\hat{S}_n] \rho(t) dW_n(t) \, ,
    \end{split}
\end{equation}
where
\begin{equation}
    \mathcal{H}[\hat{O}] \rho = \hat{O} \rho + \rho \hat{O}^\dagger - \mathrm{Tr}[\hat{O} \rho + \rho \hat{O}^\dagger] \rho \, .
\end{equation}
The equations now take into account also the presence of additional loss channels $\hat{C}_i$ not related to the measurement processes $\hat{S}_n$. The stochastic master equation can be solved using the \code{smesolve} function, which takes the same arguments as the \code{ssesolve} function, but also requires the list of collapse operators \code{c_ops}. Here, we consider the same system as before, but we also include the qubit decay channel $\hat{C}_1 = \sqrt{\gamma} \hat{\sigma}_-$ and cavity pure dephasing $\hat{C}_2 = \sqrt{\kappa_\varphi} \hat{a}^\dagger \hat{a}$.

\begin{lstlisting}
# Including qubit losses
κφ = 0.3    # cavity dephasing rate
c_ops  = [sqrt(γ) * σm, sqrt(κφ) * a' * a]

sol_sme = smesolve(
    H,
    ψ0,
    tlist,
    c_ops,
    sc_ops[1],
    e_ops = [X],
    ntraj = ntraj,
    store_measurement = Val(true)
)
sol_sme.expect      # get expectation values
sol_sme.measurement # get homodyne current
\end{lstlisting}

As in the previous case, it is possible to extract the expectation value $\langle \hat{X} \rangle$ and the homodyne current $J_x$ from the \code{sol_sme} object. Also, a simpler algorithm specific for single Wiener process is used to solve the SME when the stochastic collapse operator is a single \code{QuantumObject}. Figure \ref{fig-sse-sme}(b) shows the time evolution of the homodyne current $J_x$ (light blue curve) and the expectation value $\langle \hat{X} \rangle$ (dark blue curve), averaged over 500 trajectories.

Support for heterodyne detection, which involves simultaneous monitoring of two orthogonal quadratures and results in a complex-valued measurement current, is planned for future releases for both the SSE and SME solvers. This will further extend the package’s capabilities in simulating realistic quantum measurement scenarios and feedback schemes.

\subsection{Dynamical solvers}
All the time-evolution solvers described above allow the use of user-defined callbacks. Indeed, thanks to the \texttt{DiffEqCallbacks.jl}, which is one of the core packages of \texttt{DifferentialEquations.jl}~\cite{DifferentialEquations2017}, it is possible to apply a given operation to the state of the system, triggered by a condition that is verified along the dynamics. To demonstrate the power of callbacks, here we show the dynamical Fock dimension (DFD) and dynamical shifted Fock (DSF) algorithms. The first is a simple extension of \code{mesolve} with an additional callback that continuously monitors the population of the Fock states in time, increasing or decreasing the cutoff dimension of the Hilbert space accordingly. In this way, it is no longer required to check for the convergence of the results as a function of a fixed cutoff in the Hilbert space dimension. This method is accessible through the \code{dfd\_mesolve}.

\begin{figure*}[t]
    \centering
    \includegraphics[width=\textwidth]{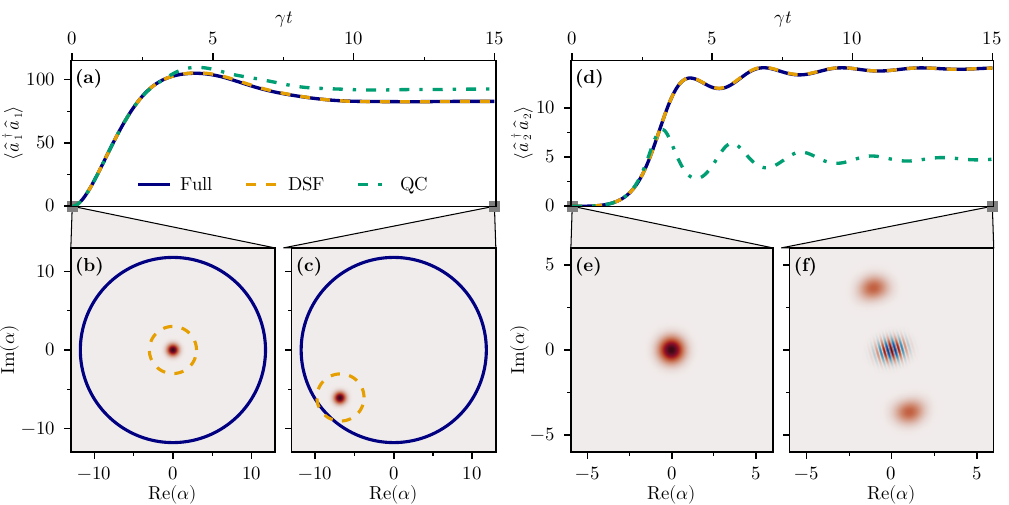}
    \caption{\textbf{Dynamical shifted Fock algorithm.} \textbf{(a)} Time evolution of the average photon number $\langle \hat{a}_1^\dagger \hat{a}_1 \rangle$ for two nonlinearly coupled harmonic oscillators with Kerr nonlinearity. In contrast to the second-order quantum cumulant expansion (dash-dotted green curve), the DSF algorithm (dashed orange curve) is able to capture the system’s dynamics while still maintaining a low-dimensional Hilbert space in the shifted Fock basis. \textbf{(b-c)} Wigner function of the first cavity field at the initial time (\textbf{(b)}), and at the final time (\textbf{(c)}). The solid blue circle represents the Hilbert space required to simulate the model using \code{mesolve}, while the dashed orange circle represents the Hilbert space used by the DSF algorithm. As can be seen, the center of the space is shifted, depending on the coherence of the state. \textbf{(d)} Average photon number of the second mode, which is simulated in the original Fock basis during the DSF algorithm solving process. However, the quantum cumulants expansion is not able to reproduce this behavior. \textbf{(e-f)} Wigner function of the second mode at the initial and final time, respectively. The Wigner function is computed using the \code{wigner} function.}
    \label{fig-dsf}
\end{figure*}

Here, we focus on the DSF method, which is a simple but powerful algorithm that allows the simulation of strongly-driven systems. Indeed, the numerical simulation of strongly-driven, nonlinear, and coupled systems presents several challenges. On one hand, the dynamics of such systems may necessitate the description of states that occupy a significant portion of the Hilbert space, making them difficult to simulate using standard numerical techniques. On the other hand, although semiclassical approximations (e.g., mean-field or cumulant expansion~\cite{Kubo1962Generalized,Plankensteiner2022QuantumCumulants_jl,Polkovnikov2010Phase}) partially address this issue, they may fail to accurately characterize quantum fluctuations. The DSF is a numerical algorithm designed to efficiently tackle these challenges. A detailed description of the algorithm is provided in Appendix~\ref{app:DSF-algorithm}, while here we limit ourselves to a brief overview of the working principle. We apply DSF to the Lindblad master equation, but the same approach can also be used to integrate Monte-Carlo quantum trajectories. 

The DSF algorithm efficiently simulates strongly driven systems by maintaining a low-dimensional Hilbert space. This is achieved through monitoring the coherence $\alpha$ and applying unitary transformations once a specified threshold value of the coherence is reached. This approach can be seen as a continuous shift of the origin of the phase space, which is continuously re-centered at $\alpha$. In order to demonstrate the full potential of the DSF algorithm, we consider a system of two harmonic oscillators with a nonlinear interaction term, a Kerr nonlinearity, and both one- and two-photon losses. More specifically, the Hamiltonian reads
\begin{equation}
    \begin{split}
            \hat{H} = & \, \Delta_1 \hat{a}_1^\dagger \hat{a}_1 + \Delta_2 \hat{a}_2^\dagger \hat{a}_2 - U \hat{a}_2^{\dagger 2} \hat{a}_2^2 \\
            &+ J \big(\hat{a}_1 \hat{a}_2^{\dagger 2} + \hat{a}_1^\dagger \hat{a}_2^2\big) + F (\hat{a}_1 + \hat{a}_1^\dagger) \, ,
    \end{split}
\end{equation}
with dissipators $\mathcal{D}[\hat{a}_1]$, $\mathcal{D}[\hat{a}_2]$, and $\mathcal{D}[\hat{a}_2^2]$. Here $\Delta_1$ and $\Delta_2$ are the detunings with respect to the drive frequency of the first and second mode, respectively; $U$ is the Kerr nonlinearity of the second mode; $J$ is the nonlinear coupling strength; and $F$ is the drive amplitude. The system evolves according to the Lindblad master equation, as shown in \cref{eq-lindblad-master-equation}.

When $J = 0$, the first mode is driven and dissipative but remains in a coherent state. For $J \neq 0$, nonlinear effects become relevant, causing the state of the first mode to deviate from a perfect coherent state while still remaining localized in phase space around it, with quantum fluctuations shaping the dynamics. Under suitable parameters, this model can generate cat states in the second mode. Such states cannot directly benefit from the DSF algorithm, since their coherence vanishes and they significantly delocalize in phase space. However, the DSF algorithm allows the user to selectively choose which subsystems should be displaced and which should be treated in the original Fock basis. This flexibility is a key advantage compared to methods such as the quantum cumulant expansion~\cite{Kubo1962Generalized}, where quantum fluctuations must be expanded for the entire system.  

The DSF algorithm is implemented in \texttt{QuantumToolbox.jl} through the function \code{dsf\_mesolve}. The user can specify a threshold for the coherence, a maximum cutoff dimension for the fluctuations, and numerical parameters such as the Krylov subspace dimension used to efficiently compute the action of the displacement operator. In order to support multipartite systems, the argument \code{op_list} must be a list of operators and each element representing the bosonic annihilation operator of a subsystem on which the DSF algorithm is applied. Because these operators evolve during the solving process, the syntax requires specifying the Hamiltonian, collapse operators, and observables as functions of \code{op_list}. For example, we demonstrate the case where we apply DSF algorithm to the first mode and the second mode remain in Fock basis.

\begin{lstlisting}
F   = 7.0
Δ1  = 0.5
Δ2  = 0.5
U   = 0.05
κ1  = 1.0
κ2  = 0.0
κ2_2 = 0.01
J   = 0.05

tlist = range(0, 15, 1000)

# Hamiltonian
function H(op_list, p)
    a1 = op_list[1]
    Δ1 * a1'*a1 + Δ2 * a2'*a2 - U * (a2^2)' * a2^2 + F * (a1 + a1') + J * (a1'*a2^2 + a1*(a2^2)')
end

# Collapse operators
function c_ops(op_list,p)
    a1 = op_list[1]
    [sqrt(κ1) * a1, sqrt(κ2) * a2, sqrt(κ2_2) * a2^2]
end

# Expectation operators
function e_ops(op_list,p)
    a1 = op_list[1]
    [a1' * a1, a2' * a2, a1, a2]
end

N1 = 4 # Cutoff of the quantum fluctuations
N2 = 40 # Cutoff of the second mode

a1 = tensor(destroy(N1), qeye(N2))
a2 = tensor(qeye(N1), destroy(N2))

# Initial state representing the quantum fluctuations
# around the initial coherent state
ψ0  = kron(fock(N1, 0), fock(N2, 0))

op_list = [a1] # Annihilation operator

sol_dsf = dsf_mesolve(H, ψ0, tlist, c_ops, op_list, e_ops = e_ops)
sol_dsf.expect # get expectation values
\end{lstlisting}
As for the other solvers, \code{sol_dsf.expect} gives the expectation values of the operators.

\begin{figure*}[t]
    \centering
    \includegraphics[width=\textwidth]{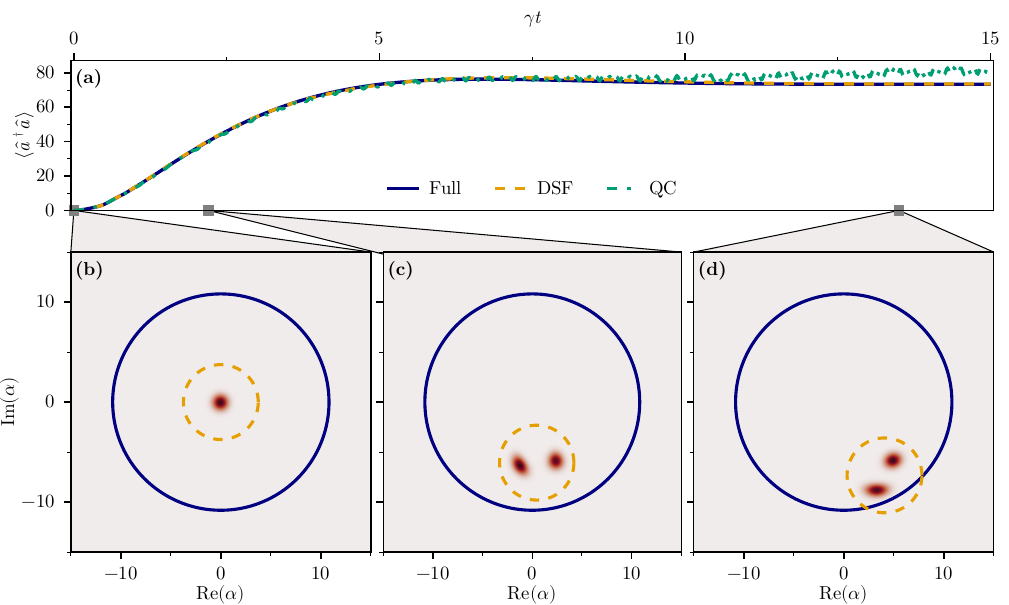}
    \caption{\textbf{Dynamical shifted Fock algorithm.} \textbf{(a)} Time evolution of the average photon number $\langle \hat{a}^\dagger \hat{a} \rangle$ for the driven Jaynes--Cummings model with Kerr nonlinearity. Even though the cavity state is not strictly coherent, its localization in phase space allows the DSF algorithm to reproduce the dynamics accurately, while the quantum cumulant expansion fails to capture the correct behavior. \textbf{(b--d)} Wigner functions of the cavity state at different times, illustrating its localized, non-coherent structure and the efficiency of the DSF representation with $N_\mathrm{dsf}=13$.}
    \label{fig-dsf-JC}
\end{figure*}

Figure~\ref{fig-dsf}(a) shows the time evolution of the average photon number $\langle \hat{a}_1^\dagger \hat{a}_1 \rangle$, computed with the standard \code{mesolve} function (solid blue curve), with the DSF algorithm (dashed orange curve), and with the second-order quantum cumulant expansion (dash-dotted green curve). The quantum cumulant expansion is a systematic method that expresses the dynamics of operator expectation values in terms of connected correlations (cumulants), truncating higher-order terms to approximate the dynamics by a closed set of lower-order equations~\cite{Kubo1962Generalized}. Here we employ \texttt{QuantumCumulants.jl}~\cite{Plankensteiner2022QuantumCumulants_jl}, which automatically generates the equations of motion and integrates them numerically.

Panels~\ref{fig-dsf}(b-c) show the Wigner function of the first mode at different times, illustrating its coherent-like shape. The solid blue circle indicates the Hilbert-space region required for the full \code{mesolve} simulation, while the dashed orange circle indicates the effective Hilbert space explored by the DSF algorithm. As expected, the displacement shifts the center of the basis, keeping the fluctuations localized. Although the first mode remains close to a coherent state, the cumulant expansion fails to reproduce the correct evolution, as nonlinear correlations induced by the second mode become important. This discrepancy is even more pronounced in Fig.~\ref{fig-dsf}(d), which shows the photon number of the second mode. Figures~\ref{fig-dsf}(e-f) display the Wigner function of the second mode, illustrating the emergence of a cat state. In this case, the DSF algorithm correctly reverts to the standard Fock basis, while the cumulant expansion completely fails to capture the non-Gaussian features.

It is worth emphasizing that the DSF algorithm does not require the state to remain strictly coherent-like. The essential condition for DSF to provide an advantage over the full Fock basis is that the quantum fluctuations remain small compared to the displacement in phase space. Larger fluctuations can still be handled, but they demand a correspondingly larger cutoff dimension in the fluctuation space. As an explicit example, we consider the Jaynes--Cummings model with a Kerr nonlinearity, driven at frequency $\omega_d$ with amplitude $F$: 
\begin{equation}
    \begin{split}
        \hat{H} = & \ \Delta_c \hat{a}^\dagger \hat{a} + \frac{\Delta_q}{2} \hat{\sigma}_z + U \hat{a}^{\dagger 2} \hat{a}^2 \\
        &+ g \big(\hat{a} \hat{\sigma}_+ + \hat{a}^\dagger \hat{\sigma}_-\big) + F (\hat{a} + \hat{a}^\dagger) \, ,
    \end{split}
\end{equation}
where $\Delta_c = \omega_c - \omega_d$ and $\Delta_q = \omega_q - \omega_d$ are the detunings of the cavity and qubit with respect to the drive frequency, and $U$ is the Kerr nonlinearity.  

Figure~\ref{fig-dsf-JC}(a) displays the time evolution of the cavity photon number, showing that the DSF algorithm reproduces the exact dynamics with high accuracy, while the quantum cumulant expansion deviates from the true case. Figure~\ref{fig-dsf-JC}(b-d) show the corresponding Wigner functions at different times. In this case the state is not coherent, but it remains localized in phase space, requiring only a modest cutoff of $N_\mathrm{dsf}=13$ for the fluctuation space.

The DSF algorithm can be also extended to the quantum trajectories method. This can have some benefits in the presence of bistable states, where the density matrix is too spread out in the Hilbert space, but the single trajectory is localized. It can be used with the \code{dsf\_mcsolve} function, with a very similar syntax.

\section{GPU and distributed computing}
\label{sec-gpu-distributed}

In recent years, graphics processing units (GPUs) have become essential tools for scientific computing due to their massive parallelism and high memory bandwidth. Originally developed for graphics rendering, GPUs gained widespread attention with the rise of deep learning~\cite{Krizhevsky2012ImageNet,LeCun2015Deep}, where they dramatically accelerated training times for large neural networks. This success sparked a broader interest across various scientific domains, where parallel numerical tasks can benefit from GPU acceleration.

The field of quantum simulation is no exception. As the complexity of quantum systems scales exponentially with system size, GPU computing offers a powerful way to tackle large Hilbert spaces and intensive linear algebra operations. This has led to a growing number of quantum toolkits supporting GPU backends to enable faster and more scalable simulations.

As already mentioned in \cref{sec-package-architecture}, the \code{QuantumObject} constructor supports any general \code{AbstractArray} type, including GPU arrays. This means that the user can easily create a \code{QuantumObject} on the GPU by passing a GPU array to the constructor. Moreover, \texttt{QuantumToolbox.jl} has special functions that allows to convert a \code{QuantumObject} from the CPU to the GPU and vice versa. This is particularly useful when the user wants to perform some operations on the CPU and then transfer the result to the GPU for further processing. Here we show how easily this can be done in the case of CUDA arrays, by applying the \code{cu} function to a generic \code{QuantumObject}. For instance, the following code runs the master equation solver on the GPU:
\begin{lstlisting}
using QuantumToolbox
using CUDA

N = 20 # cutoff of the Hilbert space dimension
ω = 1.0 # frequency of the harmonic oscillator
γ = 0.1 # damping rate

tlist = range(0, 10, 100) # time list

# The only difference in the code is the cu() function
a = cu(destroy(N))

H_gpu = ω * a' * a

ψ0_gpu = cu(fock(N, 3))

c_ops = [sqrt(γ) * a]
e_ops = [a' * a]

sol = mesolve(H_gpu, ψ0_gpu, tlist, c_ops, e_ops = e_ops)
sol.expect # get expectation values
\end{lstlisting}
This can also be applied to many other functions. For example, the computation of the Wigner function can become computationally demanding in the cases of a large Hilbert space or of a fine grid in phase space. By converting the phase space grid to a GPU array, the computation can be significantly accelerated.
\begin{lstlisting}
N = 200
n_ph = 100
α = sqrt(n_ph)

ψ = normalize(coherent(N, α) + coherent(N, -α))

xvec = CuVector(range(-15, 15, 500))
yvec = CuVector(range(-15, 15, 500))

wig = wigner(ψ, xvec, yvec)
\end{lstlisting}
This can be very useful when rendering animations of the time evolution of the Wigner function, reducing the time needed to compute each frame. A performance comparison between the CPU and GPU implementations together with the comparison with other packages will be shown in \cref{sec-performance-comparison}.

\texttt{QuantumToolbox.jl} also supports distributed computing using Julia’s built-in \texttt{Distributed.jl} package. This allows users to parallelize simulations across multiple CPU cores or nodes, which is especially useful for parameter sweeps, Monte Carlo trajectories, or ensemble averages. The framework is designed to distribute workloads efficiently while keeping the user interface simple and intuitive. Combined with GPU acceleration, this enables hybrid parallelism to tackle large and computationally demanding quantum simulations.

We now demonstrate distributed computing in the context of embarrassingly parallel problems, such as Monte Carlo trajectories and ensemble simulations. Thanks to Julia’s built-in \texttt{Distributed.jl} package, these tasks can be parallelized with minimal effort. We also note that \texttt{QuantumToolbox.jl} is fully compatible with Julia’s \code{AbstractArray} interface, which opens the possibility of using distributed array backends such as \texttt{DistributedArrays.jl}~\cite{DistributedArrays_jl}, \texttt{PartitionedArrays.jl}~\cite{PartitionedArrays_jl}, or \texttt{Dagger.jl}~\cite{Dagger_jl} for more sophisticated distributed linear algebra, although this functionality is beyond the scope of the present demonstration.

Let us now consider the two-dimensional transverse field Ising model on a $4 \times 4$ lattice. The Hamiltonian is given by
\begin{equation}
    \hat{H} = J_z \sum_{\langle i,j \rangle} \hat{\sigma}_i^z \hat{\sigma}_j^z + h_x \sum_i \hat{\sigma}_i^x \, ,
\end{equation}
where the sums are over nearest neighbors. We now study the evolution of the system using the quantum Monte Carlo method, under the local dissipators $\hat{c}_i = \sqrt{\gamma} \hat{\sigma}_i^-$, where $\gamma$ is the decay rate. Quantum trajectories are very useful here, as they avoid the need to compute the full Liouvillian of such a large system. Moreover, they can be seamlessly parallelized by simply adding the \code{ensemblealg} keyword argument to the \code{mcsolve} function. Indeed, thanks to the Distributed.jl and SlurmClusterManager.jl packages, it is possible to parallelize on a cluster with minimal effort. The following examples are applied to a cluster with the SLURM workload manager, but the same principles can be applied to other workload managers. The script file simulating the dynamics of the system is given by
\begin{lstlisting}
using Distributed
using SlurmClusterManager

const SLURM_CPUS_PER_TASK = get(ENV, "SLURM_CPUS_PER_TASK", 1)

exeflags = ["--project=.", "-t $SLURM_CPUS_PER_TASK"]
addprocs(SlurmManager(); exeflags=exeflags)

@everywhere begin
    using QuantumToolbox
    import SciMLBase: EnsembleSplitThreads

    BLAS.set_num_threads(1)
end

# Define lattice
Nx = 4
Ny = 4
latt = Lattice(Nx = Nx, Ny = Ny)

# Define Hamiltonian and collapse operators
Jx = 0.0
Jy = 0.0
Jz = 1.0
hx = 0.2
hy = 0.0
hz = 0.0
γ = 1

Sx = mapreduce(i->multisite_operator(latt, i=>sigmax()), +, 1:latt.N)
Sy = mapreduce(i->multisite_operator(latt, i=>sigmay()), +, 1:latt.N)
Sz = mapreduce(i->multisite_operator(latt, i=>sigmaz()), +, 1:latt.N)

H, c_ops = DissipativeIsing(Jx, Jy, Jz, hx, hy, hz, γ, latt; boundary_condition = Val(:periodic_bc), order = 1)
e_ops = [Sx, Sy, Sz]

# Time Evolution

ψ0 = fock(2^latt.N, 0, dims = ntuple(i->2, Val(latt.N)))

tlist = range(0, 10, 100)

sol_mc = mcsolve(H, ψ0, tlist, c_ops, e_ops=e_ops, ntraj=5000, ensemblealg=EnsembleSplitThreads())

rmprocs(workers())

sol_mc.expect # get expectation values
\end{lstlisting}
The average expectation value $\sum_i\langle\hat{\sigma}^x_i\rangle$, $\sum_i\langle\hat{\sigma}^y_i\rangle$, and $\sum_i\langle\hat{\sigma}^z_i\rangle$ at each time point in \code{tlist} can be extracted by \code{sol_mc.expect[1,:]}, \code{sol_mc.expect[2,:]}, and \code{sol_mc.expect[3,:]}, respectively.

Here, we used some helper functions to easily define the Hamiltonian and the collapse operators. The simulation of 5000 trajectories took 6 minutes on 10 nodes with 72 threads each, for a total of 720 threads on a Intel(R) Xeon(R) Platinum 8360Y CPU @ 2.40 GHz processor.

\section{Automatic differentiation}
\label{sec-automatic-differentiation}

In many applications, one is interested not only in simulating the dynamics of an open quantum system, but also in optimizing some cost functional that depends on the system’s evolution. Typical examples include quantum optimal control, variational ansätze for open dynamics, or parameter estimation~\cite{Peirce1988Optimal,Werschnik2007Quantum,D_Alessandro2021Introduction,Koch2022Quantum}. These tasks require the efficient computation of derivatives of observables with respect to physical parameters.

Let us consider a generic quantum system described by a time- and parameter-dependent Hamiltonian $\hat{H}(t,\mathbf{p})$, and a set of collapse operators $\hat{C}_k(t,\mathbf{p})$, where $\mathbf{p} = (p_1,\dots,p_m)$ denotes a vector of parameters. The dynamics of the density matrix $\hat{\rho}(t,\mathbf{p})$ is governed by the Lindblad master equation in \cref{eq-lindblad-master-equation}, namely
\begin{equation}
\begin{split}
    \frac{d}{dt} \hat{\rho} (t,\mathbf{p})
    =& -i\big[\hat{H}(t,\mathbf{p}),\,  \hat{\rho} (t,\mathbf{p})\big] \\
    &+ \sum_k \mathcal{D} \big[ \hat{C}_k(t,\mathbf{p}) \big]  \hat{\rho} (t,\mathbf{p}) \; .
\end{split}
\label{eq:lindblad_ad}
\end{equation}
The expectation value of an observable $\hat{O}$ at a given time $T$ is then
\begin{equation}
f(\mathbf{p}) = \langle \hat{O} \rangle (T, \mathbf{p}) 
= \mathrm{Tr}\!\left[ \hat{O} \, \hat{\rho}(T,\mathbf{p}) \right].
\end{equation}
In many practical scenarios, one is interested in the gradient
\begin{equation}
\nabla_{\mathbf{p}} f(\mathbf{p}) = 
\left( \frac{\partial f}{\partial p_1}, \cdots, \frac{\partial f}{\partial p_m} \right),
\end{equation}
which quantifies how the final observable depends on the system’s parameters. This gradient is essential in optimization and control, where $f(\mathbf{p})$ is used as a cost function. A straightforward approach is to approximate these derivatives using finite differences. However, this is computationally expensive, requiring at least $m+1$ full simulations, and numerically unstable, especially when $f(\mathbf{p})$ results from the integration of stiff differential equations.

Automatic differentiation (AD) provides an efficient alternative. It computes derivatives of $f(\mathbf{p})$, with respect to all parameters $\mathbf{p}$, to machine precision by applying the chain rule directly to the solver’s sequence of elementary operations.

In \texttt{QuantumToolbox.jl}, we have introduced preliminary support for automatic differentiation. Many of the core functions are compatible with AD engines such as \texttt{ForwardDiff.jl}~\cite{ForwardDiff_jl}, \texttt{Zygote.jl}~\cite{Zygote_jl2018} and \texttt{Enzyme.jl}~\cite{Enzyme_jl2020,Enzyme_jl2021,Enzyme_jl2022}, allowing users to compute gradients of observables or cost functionals involving the time evolution of open quantum systems. Both forward and reverse AD are supported, in particular:
\begin{itemize}
    \item \textbf{Forward-mode AD} (via \texttt{ForwardDiff.jl} or \texttt{Enzyme.jl}): propagates derivatives of each parameter alongside the system state during time evolution. Its cost scales linearly with the number of parameters $m$, making it efficient when $m$ is small (for example, when optimizing only a few drive amplitudes or detunings).
    \item \textbf{Reverse-mode AD} (via \texttt{Zygote.jl} or \texttt{Enzyme.jl}): propagates sensitivities backward from the output quantity to all parameters at once. The cost scales with the number of outputs rather than the number of parameters. For scalar objectives such as $f (\mathbf{p})$ or a cost functional in optimal control, the computational cost of reverse-mode AD is essentially comparable to one (or a few) primal solves, regardless of $m$. This makes it the method of choice for problems with many parameters, as in pulse-shaping or variational optimization. Reverse-mode AD corresponds to the ``backpropagation'' approach widely used in machine learning.
\end{itemize}

Although \texttt{QuantumToolbox.jl} was not originally designed with AD in mind, its architecture facilitated the integration of AD capabilities. Many core functions were already compatible with AD engines out of the box.

At present, this functionality is considered experimental and not all parts of the library are AD-compatible. Nevertheless, these initial results demonstrate the flexibility of the \texttt{QuantumToolbox.jl} design and open the door to advanced applications, such as quantum optimal control and variational optimization, within the same simulation framework. Here we provide a simple example of how to use the reverse-mode AD capabilities of \texttt{QuantumToolbox.jl}, by computing the gradient of a master equation time evolution with respect to some Hamiltonian parameters.

\begin{figure}[t]
    \centering
    \includegraphics{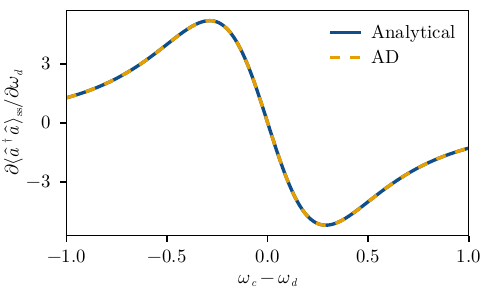}
    \caption{\textbf{Automatic differentiation.} Derivative of the expectation value $\langle \hat{a}^\dagger \hat{a} \rangle$ at the steady state with respect to the drive frequency $\omega_d$. The solid blue curve represents the analytical derivative, while the dashed orange curve represents the derivative computed using automatic differentiation, AD.}
    \label{fig-autodiff}
\end{figure}

Let us consider the case of a driven-dissipative quantum harmonic oscillator, where the Hamiltonian is given by

\begin{equation}
    \hat{H} (t) = \omega_c \hat{a}^\dagger \hat{a} + F (\hat{a} e^{i \omega_d t} + \hat{a}^\dagger e^{-i \omega_d t}) \, ,
\end{equation}
where $\omega_d$ is the drive frequency, and $F$ is the drive amplitude. The system evolves according to the Lindblad master equation in \cref{eq-lindblad-master-equation}, where the collapse operator is given by $\hat{C} = \sqrt{\gamma} \hat{a}$, with $\gamma$ being the decay rate. The time evolution of the system can be computed analytically, and the number of photons at the steady state is given by
\begin{equation}
    \langle \hat{a}^\dagger \hat{a} \rangle_\mathrm{ss} = \frac{F^2}{(\omega_c - \omega_d)^2 + \frac{\gamma^2}{4}} \, ,
\end{equation}
and its derivative with respect to the drive frequency is given by
\begin{equation}
    \frac{\partial \langle \hat{a}^\dagger \hat{a} \rangle_\mathrm{ss}}{\partial \omega_d} = \frac{2 F^2 (\omega_c - \omega_d)}{\big[ (\omega_c - \omega_d)^2 + \frac{\gamma^2}{4} \big]^2} \, .
\end{equation}
Thanks to the \texttt{SciMLSensitivity.jl}~\cite{SciMLSensitivity_jl} package, which is part of the \texttt{DifferentialEquations.jl} ecosystem, we can compute the derivative of the ordinary differential equation (ODE) using various algorithms. Here we use the adjoint sensitivity algorithm using a backwards solution of the ODE to compute the gradient with respect to $\omega_d$ and $F$.
\begin{lstlisting}
using SciMLSensitivity
using Zygote

N  = 20
a  = destroy(N)
ωc = 5.0
γ  = 1.0

coef_a(p, t) = p[2] * exp(1im * p[1] * t)
coef_ac(p, t) = conj(coef_a(p, t))

H = ωc * a' * a + QobjEvo(a, coef_a) + QobjEvo(a', coef_ac)
c_ops = [sqrt(γ) * a]
L = liouvillian(H, c_ops)
ψ0 = fock(N, 0)
tlist = range(0, 40, 100)

function my_f_mesolve(p)
    sol = mesolve(
        L,
        ψ0,
        tlist,
        progress_bar = Val(false),
        params = p,
        sensealg = BacksolveAdjoint(autojacvec = EnzymeVJP()),
    )

    return real(expect(a' * a, sol.states[end]))
end

ωd = 1.0
F = 1.0
params = [ωd, F]

grad = Zygote.gradient(my_f_mesolve, params)[1]
\end{lstlisting}

Once again, the syntax is very simple, as the only difference is the addition of the \code{params} and \code{sensealg} keyword arguments to the \code{mesolve} function. The parameters are passed as a \code{Vector}, and the dependence of the Hamiltonian on the parameters is defined using the \code{QobjEvo} constructor. Figure~\ref{fig-autodiff} shows the comparison between the analytical derivative (solid blue curve) and the derivative computed using automatic differentiation (dashed orange curve). As can be seen, the two curves are in perfect agreement, demonstrating the accuracy of the AD implementation.

The previous example illustrates how gradients can be obtained by back-propagating a master equation evolution. Reverse-mode AD becomes especially powerful when optimizing over many parameters, such as in pulse-shaping tasks aimed at preparing a desired target state. As a demonstration, we consider a two-qubit system initialized in $\ket{0,0}$, with the goal of generating the Bell state $\ket{\Phi^+} = (\ket{0,0} + \ket{1,1})/\sqrt{2}$ at a final time $t_f$.

In the absence of losses, this state can be ideally prepared by applying a Hadamard gate on the first qubit, followed by a CNOT gate~\cite{Nielsen-Chuang2011}. This corresponds to the time-dependent Hamiltonian
\begin{equation}
    \hat{H}_\mathrm{Bell}(t) = \hat{H}^{(1)}_\mathrm{H}(t) + \hat{H}_\mathrm{CNOT}(t),
\end{equation}
with
\begin{equation}
    \hat{H}^{(1)}_\mathrm{H}(t) = f_\mathrm{H}(t) \, \frac{\pi}{t_f \sqrt{2}} \left( \hat{\sigma}_x^{(1)} - \hat{\sigma}_z^{(1)} \right),
\end{equation}
representing the Hadamard rotation on qubit 1, active during the first half of the evolution through $f_\mathrm{H}(t) = \Theta(t)\Theta(t_f/2 - t)$, where $\Theta$ is the Heaviside step function, and
\begin{equation}
    \hat{H}_\mathrm{CNOT}(t) = f_\mathrm{CNOT}(t) \, \frac{\pi}{t_f} \left(\hat{I}^{(1)} + \hat{\sigma}_z^{(1)} \right) \otimes \left( \hat{I}^{(2)} - \hat{\sigma}_x^{(2)} \right),
\end{equation}
implementing the CNOT gate during the second half of the protocol, with $f_\mathrm{CNOT}(t) = \Theta(t - t_f/2)\Theta(t_f - t)$.

When including dissipation via collapse operators $\hat{C}_j = \sqrt{\gamma}\,\hat{\sigma}_-^{(j)}$ with $\gamma = 10^{-2}$ and $t_f = 100$, the resulting fidelity is $\mathcal{F} = \mel{\Phi^+}{\hat{\rho}(t_f)}{\Phi^+} \approx 0.77$, clearly below the ideal value. We emphasize that the chosen decay rate $\gamma$ is deliberately large compared to the gate time, in order to highlight a noticeable drop in fidelity. In standard experimental architectures, gate times are typically much shorter than decoherence times.

We now assume the functional forms of $f_\mathrm{H}(t)$ and $f_\mathrm{CNOT}(t)$ are unknown, and instead optimize their shapes directly through gradient-based algorithms.

For this purpose, each function is discretized into $N$ piecewise-constant segments, with amplitudes treated as free parameters:
\begin{align}
    \tilde{f}_\mathrm{H,opt}(t) &= \sum_{i=1}^N \theta_i^\mathrm{H}\,\chi_i(t), \label{eq-piecewise-constant-hadamard}\\
    \tilde{f}_\mathrm{CNOT,opt}(t) &= \sum_{i=1}^N \theta_i^\mathrm{CNOT}\,\chi_i(t), \label{eq-piecewise-constant-cnot}
\end{align}
where $\chi_i(t)$ is unity in the interval $t_i < t < t_{i+1}$ and zero otherwise. To ensure smoothness, the final driving fields $f_\mathrm{H,opt}(t)$ and $f_\mathrm{CNOT,opt}(t)$ are obtained by cubic interpolation of the piecewise-constant functions.

\begin{figure}[t]
    \centering
    \includegraphics{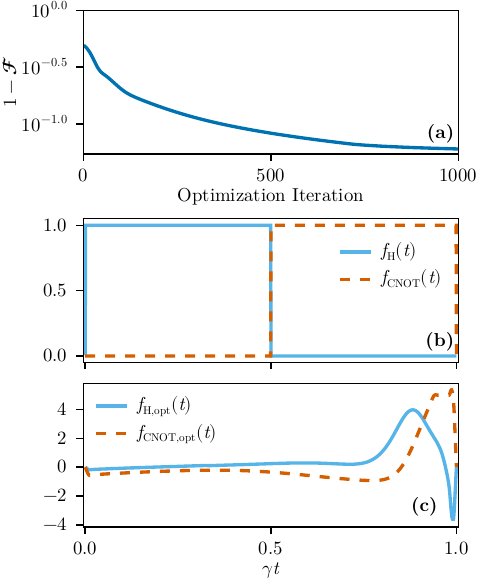}
    \caption{\textbf{Bell-state preparation via optimal control.} \textbf{(a)} Infidelity as a function of optimization steps, converging to a final fidelity $\mathcal{F}_\mathrm{opt} \approx 0.94$. \textbf{(b)} Ideal pulse shapes for the Hadamard and CNOT sequence. \textbf{(c)} Optimized pulse shapes obtained through cubic interpolation of piecewise-constant controls.}
    \label{fig:autodiff-bell}
\end{figure}

The optimization thus reduces to finding the parameter vectors $\boldsymbol{\theta}^\mathrm{H}$ and $\boldsymbol{\theta}^\mathrm{CNOT}$ that minimize the objective function, i.e., the infidelity $1-\mathcal{F}$. Using $N=100$ points per pulse, we optimize over 200 parameters $\{\theta_i^\mathrm{H},\theta_i^\mathrm{CNOT}\}_i$ with the Adam algorithm~\cite{Kingma2017Adam}. Figure~\ref{fig:autodiff-bell}(a) shows the infidelity as a function of optimization steps, converging to $\mathcal{F}_\mathrm{opt} \approx 0.94$, clearly above the standard gate sequence. To avoid unphysical values for the amplitudes, we also introduce the boundaries from $-5$ to $5$, higher values would give higher fidelities. Figure~\ref{fig:autodiff-bell}(b) illustrates the standard pulse shapes, while Fig.~\ref{fig:autodiff-bell}(c) displays the optimized ones after interpolation.

This example highlights how \texttt{QuantumToolbox.jl} leverages automatic differentiation for quantum optimal control, enabling the optimization of pulse shapes involving hundreds of parameters with minimal user effort.

\begin{figure*}[t]
    \centering
    \includegraphics[width=\textwidth]{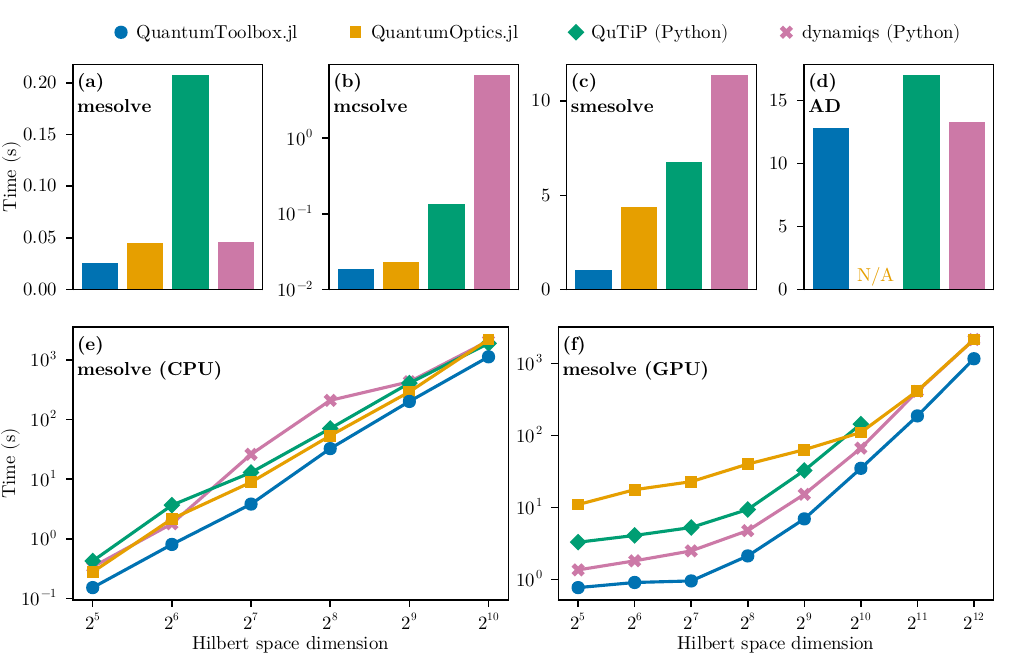}
    \caption{\textbf{Comparison of the performance with other packages.} \textbf{(a)} Master equation. \textbf{(b)} Monte Carlo quantum trajectories. \textbf{(c)} Stochastic master equation. \textbf{(d)} AD of the master equation. \textbf{(e-f)} Scaling of the master equation computation time on the CPU (\textbf{e}) and GPU (\textbf{f}) with respect to the Hilbert space dimension. \texttt{QuantumToolbox.jl} outperforms other packages in all scenarios. While the AD support in \texttt{QuantumToolbox.jl} should still be considered experimental, it is already competitive for backpropagation of the master equation. The benchmarks were performed on a workstation with an Intel(R) Core(TM) i9-13900KF CPU and 64 GB of RAM. The GPU simulations were performed on a NVIDIA GeForce RTX 4090 GPU. The versions of the packages used in the benchmarks are summarized in \cref{tab:package-versions}.}
    \label{fig:performance-comparison}
\end{figure*}

\section{Performance comparison with other packages}

\label{sec-performance-comparison}

To assess the computational efficiency of \texttt{QuantumToolbox.jl}, we benchmarked its performance against leading quantum simulation packages: \texttt{QuTiP} (Python)~\cite{QuTiP2012,QuTiP2013,QuTiP2022,QuTiP2024}, \texttt{QuantumToolbox.jl} (Julia)~\cite{QuantumOptics2018}, and \texttt{dynamiqs} (Python)~\cite{dynamiqs2024}. The benchmarks include different types of solvers (mesolve, mcsolve, and smesolve), and a range of Hilbert space dimensions, both on CPU and GPU. The code used to generate these benchmarks is available in a dedicated repository~\cite{QuantumToolbox_jl_Figures}.

In \cref{fig:performance-comparison}(a-d), we simulate the open evolution of the driven Kerr oscillator, with the Hamiltonian
\begin{equation}
    \hat{H} = \Delta \hat{a}^\dagger \hat{a} - U \hat{a}^{\dagger 2} \hat{a}^2 + F (\hat{a} + \hat{a}^\dagger)
\end{equation}
and single photon loss with rate $\gamma$. In \cref{fig:performance-comparison}(e-f), we simulate the 1D Ising model
\begin{equation}
    \hat{H} = h_z \sum_{k=1}^{N} \hat{\sigma}_k^z + J_x \sum_{j=1}^{N-1} \hat{\sigma}_k^x \hat{\sigma}_{k+1}^x
\end{equation}
with local dissipators $\{ \hat{\sigma}_j^- \}$, and we vary the number of spins $N$ up to 12 in the GPU case.

As shown in Fig.~\ref{fig:performance-comparison}, \texttt{QuantumToolbox.jl} consistently outperforms the other libraries across all tested scenarios. Notably, its GPU implementation achieves significant speedups already from small Hilbert sizes, demonstrating superior scalability. These results highlight the advantage of Julia’s performance-oriented design combined with QuantumToolbox’s efficient solver architecture.

The benchmarks were done on a workstation with an Intel(R) Core(TM) i9-13900KF CPU and 64 GB of RAM. The GPU simulations were performed on a NVIDIA GeForce RTX 4090 GPU. The GPU version of \texttt{QuTiP} was performed using its JAX backend, i.e., \texttt{qutip-jax}. Other information about the packages used in the benchmarks is summarized in \cref{tab:package-versions}.

\begin{table}[!h]
    \centering
    \small
    \begin{tabular}{cc|cc}
        \hline
        \hline
        Package & Version & Package & Version\\
        \hline
        Julia & 1.11.6 & Python & 3.13.5\\
        \texttt{QuantumToolbox.jl} & 0.34.1 & \texttt{QuTiP} & 5.2.1\\
        && \texttt{QuTiP-JAX} & 0.1.1\\
        \texttt{QuantumOptics.jl} & 1.2.3 & \texttt{dynamiqs} & 0.3.3\\
        \hline
        \hline
    \end{tabular}
    \caption{\textbf{Versions of the packages used to generate all the figures.} We list the latest version of each package as of Aug. 28, 2025.}
    \label{tab:package-versions}
\end{table}

\section{Conclusions}

We presented \texttt{QuantumToolbox.jl}, a high-performance Julia package for simulating open quantum systems. By leveraging Julia’s JIT compilation and multiple dispatch, the package achieves performance on par with or exceeding that of established tools like \texttt{QuTiP} and \texttt{QuantumToolbox.jl}, particularly in time evolution and master equation solvers. Its modular design and native compatibility with GPU computing, automatic differentiation, and advanced visualization tools make it a flexible and efficient framework for both research and teaching. The package reached its first stable release, but we plan to add new features in the future.

It is worth mentioning that, while Julia provides substantial performance advantages, it still suffers from a relatively young ecosystem, longer initial compilation times, and limited community support compared to Python. Nonetheless, we believe that Julia’s rapid evolution and growing adoption in scientific computing make it a strong foundation for building next-generation quantum simulation tools. We hope that \texttt{QuantumToolbox.jl} will contribute to this ecosystem and become a useful resource for researchers and educators working in quantum science.

\section*{Acknowledgements}
We acknowledge Fabrizio Minganti and Luca Gravina for the useful discussions, especially for the dynamical shifted Fock algorithm. We also acknowledge Filippo Ferrari, Lorenzo Fioroni, Po-Chen Kuo, Jhen-Dong Lin, Po-Rong Lai, and Hsiang-Wei Huang for their support and beta testing in the development of the package. We thank all the contributors to the package, as well as the \texttt{QuTiP} team for their support and the useful discussions. In particular, we thank Eric Gigu\`{e}re and Neill Lambert for their insightful discussions on the benchmarks.
YTH acknowledges the support of the National Science and Technology Council, Taiwan (NSTC Grant No. 113-2917-I-006-024).
YNC acknowledges the support of the National Center for Theoretical Sciences and the National Science and Technology Council, Taiwan (NSTC Grant No. 113-2123-M-006-001 and 114-2112-M-006 -015 -MY3).
V.S. acknowledges the Swiss National Science Foundation through Projects No. 200020\_215172, 200021-227992, and 20QU-1\_215928.
F.N. is supported in part by: the Japan Science and Technology Agency (JST) [via the CREST Quantum Frontiers program Grant No. JPMJCR24I2, the Quantum Leap Flagship Program (Q-LEAP), and the Moonshot R\&D Grant Number JPMJMS2061], and the Office of Naval Research (ONR) Global (via Grant No. N62909-23-1-2074).

\appendix

\section{Solving the steady state of time-dependent systems}\label{app:steadystate-for-TD-systems}
In this section we explain one the algorithms behing the \code{steadystate_fourier} function, used for extracting the time-averaged steady state for time-dependent systems. We first define the time-dependent equation of motion for the density matrix
\begin{equation}
    \frac{ d}{ d t} \hat{\rho} = \left[ \mathcal{L}_0 + \mathcal{L}_1 e^{i \omega_\mathrm{d} t} + \mathcal{L}_{-1} e^{-i \omega_\mathrm{d} t} \right] \hat{\rho} \, ,
\end{equation}
where $\mathcal{L}_0$ is the time-independent Liouvillian superoperator, while $\mathcal{L}_{\pm 1}$ are the superoperators containing the drive terms. For example, in the case of the optomechanical system in \cref{eq-optomechanical-driven}, the Liouvillian superoperators are given by
\begin{align}
    \mathcal{L}_0 \hat{\rho} = & \ -i \left[ \hat{H}, \hat{\rho} \right] + \kappa \mathcal{D}[\hat{a}] \hat{\rho} + \gamma \mathcal{D}[\hat{b}] \hat{\rho}, \nonumber \\
    \mathcal{L}_{\pm 1} \hat{\rho} = & -i \left[ \frac{F}{2} \left( \hat{a} + \hat{a}^\dagger \right), \hat{\rho} \right].
\end{align}
At long times, all the transient dynamics are washed out, and the density matrix can be expanded in Fourier components of the form
\begin{equation}
    \hat{\rho} (t) = \sum_{n=-\infty}^{+\infty} \hat{\rho}_n e^{i n \omega_\mathrm{d} t} \, .
\end{equation}
By substituting the expansion in the equation of motion, we obtain
\begin{equation}
\begin{split}
    \sum_{n=-\infty}^{+\infty} & i n \omega_\mathrm{d} \hat{\rho}_n e^{i n \omega_\mathrm{d} t} = \\
    &\sum_{n=-\infty}^{+\infty} \left[ \mathcal{L}_0 + \mathcal{L}_1 e^{i \omega_\mathrm{d} t} + \mathcal{L}_{-1} e^{-i \omega_\mathrm{d} t} \right] \hat{\rho}_n e^{i n \omega_\mathrm{d} t} \, .
\end{split}
\end{equation}
By equating the coefficients of the series yields the tridiagonal recursion relation
\begin{equation}
    ( \mathcal{L}_0 - i n \omega_\mathrm{d} ) \hat{\rho}_n + \mathcal{L}_1 \hat{\rho}_{n-1} + \mathcal{L}_{-1} \hat{\rho}_{n+1} = 0 \, .
\end{equation}
After choosing a cutoff $n_\mathrm{max}$ for the Fourier components, the equation above defines a tridiagonal linear system of the form $\mathbf{A} \cdot \mathbf{b} = 0$, where
\begin{widetext}
    \begin{equation}
        \mathbf{A} = \begin{pmatrix}
        \mathcal{L}_0 - i (-n_\mathrm{max}) \omega_\mathrm{d} & \mathcal{L}_{-1} & 0 & \cdots & 0 \\
        \mathcal{L}_1 & \mathcal{L}_0 - i (-n_\mathrm{max}+1) \omega_\mathrm{d} & \mathcal{L}_{-1} & \cdots & 0 \\
        0 & \mathcal{L}_1 & \mathcal{L}_0 - i (-n_\mathrm{max}+2) \omega_\mathrm{d} & \cdots & 0 \\
        \vdots & \vdots & \vdots & \ddots & \vdots \\
        0 & 0 & 0 & \cdots & \mathcal{L}_0 - i n_\mathrm{max} \omega_\mathrm{d}
        \end{pmatrix}
    \end{equation}
\end{widetext}
and
\begin{equation}
    \mathbf{b} = \begin{pmatrix}
    \hat{\rho}_{-n_\mathrm{max}} &
    \dots &
    \hat{\rho}_{0} &
    \dots &
    \hat{\rho}_{n_\mathrm{max}}
    \end{pmatrix}^{\top} \, .
\end{equation}
This allows to simultaneously obtain all the Fourier components $\hat{\rho}_n$.

\section{The dynamical shifted Fock algorithm}\label{app:DSF-algorithm}
The DSF algorithm is based on the efficient manipulation of coherent states and displacement operators. A coherent state is defined as:
\begin{equation}
    \label{eq-coherent-state}
    \vert \alpha \rangle = e^{-|\alpha|^2/2} \sum_{n=0}^{+ \infty} \frac{\alpha^n}{\sqrt{n!}} \vert n \rangle \, ,
\end{equation}
which can be also seen as the eigenstate of the bosonic annihilation operator $\hat{a} \vert \alpha \rangle = \alpha \vert \alpha \rangle$. We define $\alpha = \mel{\alpha}{\hat{a}}{\alpha}$ as the coherence of the state. A coherent state can be obtained from the action of the displacement operator
\begin{equation}
    \label{eq-displacement-operator}
    \hat{D} (\alpha) = e^{\alpha \hat{a}^\dagger - \alpha^* \hat{a}} \, , \quad \, \hat{D}(\alpha) \hat{D}^\dagger (\alpha) = \hat{D} (\alpha) \hat{D} (-\alpha) = 1
\end{equation}
to the vacuum state $\vert 0 \rangle$, namely $\vert \alpha \rangle = \hat{D} (\alpha) \vert 0 \rangle$. From this definition and from \cref{eq-displacement-operator}, it follows that
\begin{equation}
    \label{eq-coherent-to-vacuum}
    \vert 0 \rangle = \hat{D}^\dagger (\alpha) \vert \alpha \rangle \, .
\end{equation}
It can also be proven that, given any two coherent states $\ket{\alpha}$ and $\ket{\beta}$, one has
\begin{equation}
     \ket{\alpha} \propto \hat{D} (\alpha-\beta)  \ket{\beta}\, .
\end{equation}
The equations above are the key points of the DSF algorithm. Indeed, a coherent state defined in the Fock basis as in \cref{eq-coherent-state} can have large occupation probabilities in the high-energy states. However, thanks to \cref{eq-coherent-to-vacuum}, it is possible to rotate a high-energy coherent state into the vacuum state $\vert 0 \rangle$. Numerically speaking, this allows us to treat a coherent state with a vector where all the elements are zero, except for the first one. Hence, we can describe it with a smaller vector, given that we keep track of its coherence.

This approach is also valid when the state is almost-coherent-like, with small quantum fluctuations around a macroscopic coherent state. In this case, given such a state $\vert \psi \rangle$, we have
\begin{equation}
    \hat{D}^\dagger (\alpha) \vert \psi \rangle = \sum_n c_n \vert n \rangle \, ,
\end{equation}
where $\vert c_n \vert$ rapidly decreases as $n$ increases. Once again, this state can still be represented with a small Fock basis, and the accuracy of the DSF depends on the size of this basis.

Let us start by considering a closed quantum system with initial state $\vert \psi (t_0 = 0) \rangle$, whose unitary dynamics is governed by a generic Hamiltonian $\hat{H}$. For fixed-time-step integration, the algorithm is presented in \Cref{alg-DSF}, and its extension to adaptive-time-step integrators is straightforward as \texttt{DiffEqCallbacks.jl} supports very general integration algorithms.

\begin{table}[t]
\centering
\begin{minipage}{\linewidth}
\begin{algorithmic}[1]
\State \textbf{Initialize:} state $\vert \psi(0) \rangle$, Hamiltonian $\hat{H}$, dissipators $\{ \hat{O}_n \}_n$ and threshold $\delta_\alpha^{(\mathrm{max})}$.
\State Set $t \gets 0$.
\While{\(t < t_f\)}
    \State \textbf{Evolve:} Integrate 
    \[
    i\frac{d}{dt} \vert \psi(t) \rangle = \hat{H} \vert \psi(t) \rangle
    \]
    \Statex \hspace{\algorithmicindent} from $t$ to $t+\Delta t$.
    \State \textbf{Compute coherence}: 
    \[
    \alpha_{t+\Delta t} \gets \langle \psi(t+\Delta t) \vert \hat{a} \vert \psi(t+\Delta t) \rangle.
    \]
    \State \textbf{Compute shift}: 
    \[
    \delta_\alpha \gets \alpha_{t+\Delta t} - \alpha_{t}.
    \]
    \If{$|\delta_\alpha| > \delta_\alpha^{(\mathrm{max})}$}
        \State \textbf{Displace state:} 
        \[
        \vert \psi(t+\Delta t) \rangle \gets \hat{D}(-\delta_\alpha)\vert \psi(t+\Delta t) \rangle.
        \]
        \State \textbf{Update operators:} 
        \[
        \hat{H} \gets \hat{D}(-\delta_\alpha) \hat{H} \hat{D}(\delta_\alpha).
        \]
        \[
        \hat{O}_n \gets \hat{D}(-\delta_\alpha) \hat{O}_n \hat{D}(\delta_\alpha).
        \]
    \EndIf
    \State Set $t \gets t+\Delta t$.
\EndWhile
\end{algorithmic}
\end{minipage}
\caption{Fixed-time-step DSF Algorithm}
\label{alg-DSF}
\end{table}

To summarize, the DSF algorithm allows us to simulate the dynamics of a system by keeping the Hilbert space dimension small, at the expense of tracking the coherence and performing the unitary transformations when reaching a given threshold. 


\section{Tables of functions}\label{app:table-of-functions}

\lstset{style=tablestyle}

In this subsection, we summarize the most relevant user-accessible functions in \texttt{QuantumToolbox.jl} into several tables. \cref{tab:Func-create-Qobj} summarizes the functions that can easily construct typical quantum states and operators. \cref{tab:Func-Solvers} summarizes the time evolution solvers that compute the dynamics of a quantum system according to different methods. \cref{tab:Func-LA-Utility} summarizes the standard linear algebra operator and some utility functions for analyzing the properties and physical quantities of given quantum objects. Note that the complete list of functions is shown in the \texttt{QuantumToolbox.jl} documentation website (\href{https://qutip.org/QuantumToolbox.jl/stable/}{https://qutip.org/QuantumToolbox.jl/stable/}). Furthermore, additional information about each function can be obtained by calling ``\code{?function\_name}'' in Julia.
\begin{table*}[h]
    \centering
    \small
    \begin{tabular}{cl}
        \hline
        \hline

        Function & Description\\

        \hline
        
        \multirow{2}{*}{\textit{User-defined quantum object}} & \\ 
        &\\
        
        \code{QuantumObject}\;/\;\code{Qobj} & General time-independent quantum object\\
        \code{QuantumObjectEvolution}\;/\;\code{QobjEvo} & General time-dependent quantum object\\
        
        \hline

        \multirow{2}{*}{\textit{States}} & \\ 
        &\\
        
        \code{basis}\;/\;\code{fock}\;/\;\code{fock\_dm} & Single basis (Fock) state\\
        \code{coherent} \;/\;\code{coherent\_dm} & Single-mode coherent state\\
        \code{rand\_ket}\;/\;\code{rand\_dm} & Random state\\
        \code{thermal\_dm} & Thermal state\\
        \code{maximally\_mixed\_dm} & Maximally mixed state\\
        \code{spin\_state} & Spin state\\
        \code{spin\_coherent} & Coherent spin state\\
        \code{bell\_state} & Bell state\\
        \code{singlet\_state} & Singlet state\\
        \code{triplet\_states} & Triplet states\\
        \code{w\_state} & $W$ state\\
        \code{ghz\_state} & Greenberger–Horne–Zeilinger state\\

        \hline
        
    \multirow{2}{*}{\textit{Operators}} & \\ 
        &\\
        
        \code{eye}\;/\;\code{qeye} & Identity operator\\
        \code{destroy} & Bosonic annihilation operator\\
        \code{create} & Bosonic creation operator\\
        \code{fdestroy} & Fermionic annihilation operator\\
        \code{fcreate} & Fermionic creation operator\\
        \code{projection} & Projection operator\\
        \code{displace} & Displacement operator\\
        \code{squeeze} & Single-mode squeeze operator\\
        \code{num} & Bosonic number operator\\
        \code{phase} & Single-mode Pegg-Barnett phase operator\\
        \code{position} & Position operator\\
        \code{momentum} & Momentum operator\\
        \code{rand\_unitary} & Random unitary operator\\
        \code{sigmax} & Spin-$1/2$ Pauli-$X$ operator\\
        \code{sigmay} & Spin-$1/2$ Pauli-$Y$ operator\\
        \code{sigmaz} & Spin-$1/2$ Pauli-$Z$ operator\\
        \code{sigmap} & Spin-$1/2$ Pauli ladder ($\sigma^+$) operator\\
        \code{sigmam} & Spin-$1/2$ Pauli ladder ($\sigma^-$) operator\\
        \code{jmat} & General spin-$j$ operator\\
        \code{qft} & Discrete quantum Fourier transform operator\\
        \hline
        \hline
    \end{tabular}
    \caption{\textbf{List of functions in \texttt{QuantumToolbox.jl} for constructing typical quantum states and operators.} Note that ``\code{\_dm}'' represents the state is constructed in density matrix formalism.}
    \label{tab:Func-create-Qobj}
\end{table*}

\begin{table*}[h]
    \centering
    \small
    \begin{tabular}{cl}
        \hline
        \hline

        Solver & Description\\

        \hline

        \code{sesolve} & Solve Schr\"odinger equation\\
        \code{mesolve} & Solve master equation\\
        \code{mcsolve} & Solve time evolution using Monte Carlo wave-function method\\
        \code{ssesolve} & Solve stochastic Schr\"odinger equation~\cite{Wiseman2009Quantum}\\
        \code{smesolve} & Solve stochastic master equation~\cite{Wiseman2009Quantum}\\
        \code{lr\_mesolve} & Solve low-rank master equation~\cite{gravina2024adaptive}\\
        \code{dfd\_mesolve} & Solve master equation using the dynamical Fock dimension algorithm \\
        \code{dsf\_mesolve} & Solve master equation using the dynamical shifted Fock algorithm \\
        \code{dsf\_mcsolve} & Solve Monte Carlo quantum trajectories using the dynamical shifted Fock algorithm \\
        \code{brmesolve} & Solve Bloch-Redfield master equation~\cite{Breuer2007The}\\
        \code{heomsolve} & Solve hierarchical equations of motion~\cite{HierarchicalEOM2023}\\
        \code{steadystate} & Calculate the steady state\\
        \code{steadystate\_fourier} & Calculate the steady state of a periodically driven system\\

        \hline
        \hline
    \end{tabular}
    \caption{\textbf{List of time evolution solvers in QuantumToolbox.jl.}}
    \label{tab:Func-Solvers}
\end{table*}

\begin{table*}[h]
    \centering
    \small
    \begin{tabular}{cl}
        \hline
        \hline

        Function & Description\\

        \hline

        \multirow{2}{*}{\textit{Linear algebra operations}} & \\ 
        &\\
        \code{conj} & Conjugation of a quantum object\\
        \code{transpose}\;/\;\code{trans} & Transpose of a quantum object\\
        \code{adjoint}\;/\;\code{dag}\;/\;\codesymbol{\textquotesingle} & Adjoint of a quantum object\\
        \code{partial\_transpose} & Partial transpose of a quantum object\\
        \code{dot} & Dot product between quantum objects\\
        \code{matrix\_element} & Extract single matrix element from a quantum object\\
        \code{tr} & Trace of a quantum object\\
        \code{ptrace} & Partial trace of a quantum object\\
        \code{svdvals} & Compute singular values of a quantum object\\
        \code{norm} & Compute standard vector or Schatten $p$-norm of a quantum object\\
        \code{normalize}\;/\;\code{normalize!}\;/\;\code{unit} & Normalize a quantum object\\
        \code{inv} & Compute matrix inverse of a quantum object\\
        \code{sqrt}\;/\;\code{sqrtm}\;/\;\codesymbol{$\surd$} & Compute matrix square root of a quantum object\\
        \code{log}\;/\;\code{logm} & Compute matrix logarithm of a quantum object\\
        \code{exp}\;/\;\code{expm} & Compute matrix exponential of a quantum object\\
        \code{sin}\;/\;\code{sinm} & Compute matrix sine of a quantum object\\
        \code{cos}\;/\;\code{cosm} & Compute matrix cosine of a quantum object\\
        \code{diag} & Extract diagonal elements from a quantum object\\
        \code{proj} & Projection operator of a quantum \code{Ket} state\\
        \code{permute} & Permute the tensor structure of a quantum object\\
        \code{tidyup}\;/\;\code{tidyup!} & Remove small matrix elements in a quantum object\\
        \code{kron}\;/\;\code{tensor}\;/\;\codesymbol{$\otimes$} & Kronecker (tensor) product of quantum objects\\

        \hline

        \multirow{2}{*}{\textit{Quantum object conversions}} & \\ 
        &\\
        
        \code{cu} & Convert \code{data} into CUDA (GPU) array\\
        \code{to\_dense} & Convert \code{data} into dense array\\
        \code{to\_sparse} & Convert \code{data} into sparse array\\
        \code{ket2dm} & Convert a \code{Ket} state into density matrix (\code{Operator})\\
        \code{mat2vec}\;/\;\code{operator_to_vector} & Vectorize an \code{Operator} into \code{OperatorKet}\\
        \code{vec2mat}\;/\;\code{vector_to_operator} & Reshape (un-vectorize) an \code{OperatorKet} back to \code{Operator}\\
        \code{spre} & Generate left multiplication \code{SuperOperator}\\
        \code{spost} & Generate right multiplication \code{SuperOperator}\\
        \code{sprepost} & Generate left-and-right multiplication \code{SuperOperator}\\
        \code{liouvillian} & Generate Liouvillian \code{SuperOperator}\\
        \code{lindblad\_dissipator} & Generate Lindblad dissipator (\code{SuperOperator})\\

        \hline

        \multirow{2}{*}{\textit{Utility functions}} & \\ 
        &\\
        \code{expect} & Compute expectation value of a quantum operator\\
        \code{variance} & Compute variance of a quantum operator\\
        \code{eigvals}\;/\;\code{eigenenergies} & Compute eigenvalues of a quantum object\\
        \code{eigen}\;/\;\code{eigsolve}\;/\;\code{eigenstates} & Compute eigenvectors and eigenvalues of a quantum object\\

        \code{eigsolve\_al} & The Arnoldi-Lindblad eigensolver~\cite{Minganti2022arnoldilindbladtime}\\
        
        \code{purity} & Compute purity of a quantum state\\
        \code{entropy\_vn} & Compute Von Neumann entropy~\cite{Nielsen-Chuang2011} of a quantum state\\
        \code{entropy\_linear} & Compute quantum linear entropy~\cite{Nielsen-Chuang2011} of a quantum state\\
        \code{entropy\_conditional} & Compute quantum conditional entropy~\cite{Nielsen-Chuang2011} of a quantum state\\
        \code{entropy\_relative} & Compute quantum relative entropy~\cite{Nielsen-Chuang2011} of a quantum state\\
        \code{entropy\_mutual} & Compute quantum mutual information~\cite{Nielsen-Chuang2011} of a quantum state\\
        \code{entanglement} & Compute entanglement entropy~\cite{Nielsen-Chuang2011} of a bipartite quantum state\\
        \code{concurrence} & Compute concurrence~\cite{Hill1997} of a bipartite quantum state\\
        \code{negativity} & Compute negativity~\cite{Vidal2002} of a bipartite quantum state\\
        \code{fidelity} & Compute fidelity~\cite{Nielsen-Chuang2011} between two quantum states\\
        \code{tracedist} & Compute trace distance~\cite{Nielsen-Chuang2011} between two quantum states\\
        \code{correlation\_2op\_1t} & Compute correlation function of two operators with one time coordinate\\
        \code{correlation\_2op\_2t} & Compute correlation function of two operators with two time coordinates\\
        \code{correlation\_3op\_1t} & Compute correlation function of three operators with one time coordinate\\
        \code{correlation\_3op\_2t} & Compute correlation function of three operators with two time coordinates\\
        \code{spectrum} & Compute the power spectrum of a correlation function\\
        \code{spectrum\_correlation\_fft} &  Compute the fast Fourier transform from a given correlation function data\\
        \code{wigner} & Generate the Wigner quasipropability distribution of a quantum state\\

        \hline
        \hline
    \end{tabular}
    \caption{\textbf{List of standard linear algebra and other utility functions for quantum objects in \texttt{QuantumToolbox.jl}.}}
    \label{tab:Func-LA-Utility}
\end{table*}

\lstset{style=mainstyle}

\clearpage

\bibliographystyle{quantum}
\bibliography{reference}

\end{document}